\pdfoutput=1
\documentclass[conference]{IEEEtran}
\IEEEoverridecommandlockouts

\usepackage[T1]{fontenc}
\usepackage{cite}
\usepackage{amsmath,amssymb,amsfonts}

\usepackage{amsthm}
\usepackage{algorithm}
\usepackage{algpseudocode}
\usepackage{graphicx}
\usepackage{booktabs}
\usepackage{multirow}
\usepackage{xcolor}
\usepackage{url}
\newcommand{\key}[1]{\textbf{#1}}

\newtheorem{theorem}{Theorem}
\newtheorem{proposition}{Proposition}
\newtheorem{lemma}{Lemma}
\newtheorem{corollary}{Corollary}
\theoremstyle{definition}
\newtheorem{definition}{Definition}
\theoremstyle{remark}

\newcommand{\OPT}{\mathrm{OPT}}
\newcommand{\Mcl}{\mathcal{M}_Z}
\newcommand{\E}{\mathbb{E}}
\newcommand{\Prob}{\mathbb{P}}

\begin{document}

\title{Calibrated Early Termination with Bounded\\Optimality Gaps for Constrained Vector Search}

\author{\IEEEauthorblockN{Xinyu Zhang$^{1}$, Jingze Li$^{1}$, Baihua Zheng$^{2}$, Weiwei Sun$^{1}$\IEEEauthorrefmark{1}\thanks{\IEEEauthorrefmark{1}Corresponding author: Weiwei Sun}}
\IEEEauthorblockA{$^{1}$Fudan University, Shanghai, China\quad $^{2}$Singapore Management University, Singapore\\
\{zhangxinyu, jingzeli20, wwsun\}@fudan.edu.cn, bhzheng@smu.edu.sg}}

\maketitle

\begin{abstract}
Vector search over learned embeddings is the retrieval layer of recommendation and retrieval-augmented generation, and many applications cap how many answer items one category may contribute.
Graph indexes answer a query by best-first traversal, and where it stops determines both its cost and whether its answer is exact.
We study early termination for such constrained $k$-nearest-neighbor ($k$-NN) queries, with plain $k$-NN as a special case, and seek the cheapest stopping rule whose answer is exact with probability at least $1-\varepsilon$.
Learned early-termination methods instead target average recall, which does not guarantee exact answers.
Multiple testing can calibrate any parameterized stopping rule to our target, but it cannot tell how much a better rule with the same risk and observations would save.
Answering this needs a lower bound on the optimum of a rule family too large to enumerate, valid for the query population rather than only for the sample.
Our key idea is to declare the observations a rule may use and to consider every rule monotone in them.
Risk-constrained stopping then becomes a linear program over an order polytope, whose Lagrangian minimizers form a nested chain of staircase rules computed exactly by dynamic programming.
Fixed-sequence binomial tests along this chain calibrate our rule, CaST, whose failure probability is at most $\varepsilon$ with confidence $1-\delta$.
Lagrangian weak duality bounds the optimum of the declared class from below at all risk levels at once, without a union bound over the rules.
The ratio of CaST's cost to this bound limits its distance from the best rule of the class.
On two datasets of one million items each, CaST replayed on logged searches costs at most 1.132$\times$ the class optimum for risk levels from 0.03 to 0.10, at 95\% joint confidence per setting.
Larger savings thus need a rule outside the class.
At a common risk target on held-out queries, CaST is faster than every other calibrated rule we time, which take 4.8\% to 45.5\% more mean time.
\end{abstract}


\section{Introduction}
\begin{figure}[t]
\centering
\includegraphics[width=\columnwidth]{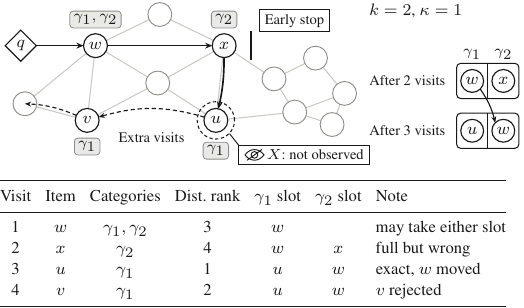}
\caption{A constrained 2-NN query with one slot per category ($\kappa=1$). An item's tag lists its categories, and each answer item fills the slot of one of them. The search visits $w$, $x$, $u$, $v$ in that order, so stopping after two visits returns $\{w,x\}$, which has $k=2$ items but is incorrect. Visiting $u$ moves $w$ to the $\gamma_2$ slot and yields the exact answer $\{u,w\}$. This happens at the completion time $X$, which the search cannot observe while it runs.
}
\label{fig:task}
\end{figure}

% Q1 what the problem is
Vector search over learned embeddings underlies recommendation and retrieval-augmented generation.
Graph indexes such as HNSW\cite{malkov2020} and DiskANN\cite{subramanya2019} answer a query by best-first traversal, iteratively exploring points based on their proximity to the query. However, because these graphs are sparsified heuristically and the search terminates once its bounded candidate list no longer improves, termination itself provides no guarantee that the exact answer has been found.
Many applications also constrain the answer.
For example, a product search should not return too many variants from the same category, and a retrieval pipeline should not draw too many passages from any single source.
We study constrained $k$-nearest-neighbor ($k$-NN) queries, in which every item carries a set of categories and the answer must admit an assignment of each answer item to one of its categories such that no
category receives more than $\kappa$ items~\cite{sonmez2022,ando2026}.
The greedy algorithm examines items in increasing order of distance to the query, selecting an item whenever the selected items, including the new one, can still be assigned to categories without violating the capacity constraints, until $k$ items are selected. To return this answer with certainty, a graph search must visit every answer item and determine when no unvisited item can change the result (Fig.~\ref{fig:task}).

% Q2 why it matters
Where the search stops thus determines the trade-off between search efficiency and answer correctness.
Our objective is a stopping rule that minimizes the expected search cost while ensuring that, for a query drawn from
the query distribution, the returned answer is exact with probability at least $1-\varepsilon$, i.e., with \emph{risk} at most $\varepsilon$.
A fixed visit budget is inefficient because the number of graph visits required to reach the exact answer, which we call the query's \emph{completion time}, varies substantially across queries.
On an HNSW graph for the Amazon (arXiv) dataset,
the budget calibrated to risk $\varepsilon=0.05$ is 6{,}781  (5{,}978) visits, i.e., 7.1$\times$ (5.8$\times$) the median completion time over 10{,}000 evaluation queries.
Learned early-termination methods, such as LAET~\cite{li2020laet} and DARTH~\cite{chatzakis2025darth}, instead predict when to stop, but they target \emph{average recall}, and  high average recall does not guarantee a high probability of returning the exact constrained answer.
On the same graphs, DARTH, configured for a target average recall of 0.95, returns the exact constrained answer for only 71\% (68\%) of queries.

% Q3 why it is hard
Finding the minimum-cost stopping rule and quantifying how close a given rule is to optimal pose three challenges.
First, the stopping time is data-dependent: the rule decides when to stop based on the search trajectory observed so far. Consequently, guarantees established at individual prediction points do not necessarily hold at the selected stopping time.
Second, the completion time is unknown at run time. Moreover, under the cap of $2^{15}$=32{,}768 visits, between 0.10\% and 0.58\% of the Amazon and arXiv queries, depending on the dataset and graph, never complete, so no rule can return their exact answer.
Third, quantifying the optimality gap requires a lower bound on the minimum expected cost over an enormous class of rules, which contains approximately $e^{266}$ deterministic rules in our setting. The lower bound must hold for the underlying query population, rather than merely for the finite evaluation sample drawn from it.

% Q4 why existing work has not solved it
Existing methods provide partial solutions to the first challenge but do not address the optimality-gap problem.
LAET~\cite{li2020laet}, DARTH~\cite{chatzakis2025darth} and DARTH+~\cite{chatzakis2026darthplus} learn stopping predictors and tune them using validation queries; DARTH+ reports recall coverage at individual prediction points, but this coverage does not necessarily extend to the data-dependent stopping point.
ConANN~\cite{horchidan2025conann} and Learn-then-Test~\cite{angelopoulos2021ltt} control the risk of a selected parameter.
Such calibration resolves the first challenge for any parameterized rule, and for a fair comparison we apply it to every baseline in our experiments.
None of these methods provides a lower bound on the minimum achievable expected cost:
a calibrated rule meets the risk target, but how much cheaper a rule with the same risk and observations could be remains unknown.
The same challenge of quantifying the optimality gap under statistical risk constraints arises in calibrated early-exit methods for language models and classifiers~\cite{schuster2022calm,jazbec2024fastyetsafe}.

% Q5 key idea and results
Our key idea is to make this question well posed by fixing what a stopping rule may observe and restricting it to a natural monotone class. At each checkpoint, our engine exposes a \emph{stopping interface}: the running minimum of a learned score (low when the answer is likely complete), discretized into buckets, together with a flag recording whether $k$ answer items have been selected. A rule is
\emph{monotone} if, whenever it stops on an observation, it also stops on every observation that indicates completion more strongly. This class contains the common threshold, cap, and budget rules aligned to the interface. With randomized rules allowed, risk-constrained stopping over the class becomes a linear program over an order polytope, whose Lagrangian minimizers form a nested chain of deterministic \emph{staircase} rules, i.e., rules whose score threshold varies monotonically across checkpoints.

This chain addresses all three challenges. First, fixed-sequence tests along the chain count failures at the actual stopping point, and thus bound the failure probability of the returned answer despite the adaptive stopping time. Second, the completion time enters only offline, through the labeled samples, i.e., sample queries whose completion times are computed from their exact answers, and the deployed rule reads nothing but the interface. Third, the same linear program yields a population lower bound on the class optimum via Lagrangian weak duality and empirical-Bernstein bounds, without a union bound over the rules of the class. 
Our technical novelty lies in exploring the structure of the monotone class to obtain this lower bound. The ratio of the calibrated cost to the lower bound is a data-dependent optimality-gap bound within the class, which holds end-to-end, from the fitted score to the returned answer. At risk $\varepsilon = 0.05$ on one graph per dataset, eight other rules calibrated to the same target spend 6.2\% to 70.6\% more visits than our calibrated staircase, which lies within $1.087\times$ of the lower bound with 87.5\% confidence.
The bound covers only rules restricted to the interface: a clairvoyant rule that knows each query's completion time is $2.8\times$ to $3.3\times$ cheaper than the empirical class optimum on the evaluation queries, which upper-bounds what richer observations could gain at these checkpoints. We evaluate in-memory search on a single machine with unfiltered traversal for queries drawn i.i.d.\ from the distribution of the labeled samples.

Our contributions are as follows.
\begin{itemize}
\item We show that risk-constrained early termination over a declared interface is a linear program over an order polytope, whose optima mix at most two nested staircases. Its Lagrangian minimizers form a nested chain computed exactly by dynamic programming in $O(J(B+1)^2)$ time per multiplier.
\item We calibrate the chain into our rule, CaST, by fixed-sequence exact binomial tests, which bound the failure probability of the returned answer by $\varepsilon$ with confidence $1-\delta$. 
We also give a lower bound on the class-optimal cost that holds simultaneously for all risk levels. Together they yield an end-to-end optimality-gap bound within the declared class.

\item On two datasets of one million items each, CaST costs at most 1.132 times the class optimum at risk levels from 0.03 to 0.10, at confidence 95\%, and this near-optimality carries over to most further sample splits, workloads and indexes. At a common risk target of 0.05 on held-out queries, CaST is faster than every other calibrated rule we time, which take 4.8\% to 45.5\% more mean time.

\end{itemize}

\section{Related Work}\label{sec:related}
\paragraph{Early termination for vector search}
LAET~\cite{li2020laet} predicts, after a fixed amount of search, how much further search a query needs and scales the prediction by a tuned factor, and Tao~\cite{yang2021tao} predicts it from static features of the query.
DARTH~\cite{chatzakis2025darth} predicts the current recall of an HNSW search with gradient-boosted trees, calls the predictor at adaptive intervals, and stops once the prediction reaches a declared target.
DARTH+~\cite{chatzakis2026darthplus} replaces the point prediction by a quantile-regression lower bound on recall, whose coverage at each prediction point is measured on test queries.
Ada-ef~\cite{zhang2026adaef} sets the list size of HNSW per query from a model of the distances between the query and the data, so that a target recall is met.
Patience in Proximity~\cite{teofili2025patience} stops once the top-$k$ answer has not changed for a number of steps.
Terminus~\cite{lu2026terminus} stops a disk-based search once recent I/Os add little utility, and E2E~\cite{xia2026e2e} learns when to terminate a filtered search.
These methods choose their stopping parameter by its average effect on validation queries.
Adaptive Beam Search~\cite{aljazzazi2025adaptive} stops on a distance-based condition that provably returns an approximate nearest neighbor on navigable graphs, which bounds the distance of the answer rather than how often it is incomplete.

\paragraph{Guarantees for approximate search}
ConANN~\cite{horchidan2025conann} uses conformal risk control~\cite{angelopoulos2024crc} to choose how many IVF clusters each query probes, so that the expected false-negative rate stays below a target.
Li et al.~\cite{li2026spanner} test a finished HNSW search and fall back to an exact recovery, Auncel~\cite{zhang2023auncel} meets a per-query error bound through an error profile, and ADSampling~\cite{gao2023adsampling} ends distance computations by hypothesis tests.
These methods control the quality of the result, and our work bounds the cost of the calibrated rule relative to the best rule of a declared class at the same risk.

\paragraph{Risk control and constrained stochastic programs}
Risk-controlling prediction sets~\cite{bates2021rcps} and Learn-then-Test~\cite{angelopoulos2021ltt} control the risk of a selected parameter, and fixed-sequence testing is one of the instruments of the latter.
Calibrated early exit applies such tests to the depth at which a language model or a classifier stops~\cite{schuster2022calm,jazbec2024fastyetsafe}.
Sample approximations of chance-constrained programs yield feasible solutions and statistical bounds on the optimal value~\cite{luedtke2008,wang2008saa}, from which Monte Carlo bounding estimates optimality gaps~\cite{mak1999}.
Our calibration and lower bound are of these kinds, and what we add is structure: an order polytope whose Lagrangian minimizers form a nested chain, which yields a lower bound along the whole cost-risk frontier.

\paragraph{Diverse and constrained nearest-neighbor search}
Query result diversification trades relevance against mutual dissimilarity~\cite{vieira2011diversification}. Diverse similarity search caps the answer items per category, with graph indexes for one category per item~\cite{anand2025} or color bucketing over black-box indexes~\cite{chen2026}.
With several categories per item, capping the number of members of each category gives a family that is not a matroid in general, and finding a feasible answer of a given size is NP-hard~\cite{celis2018}.
The assignment constraint we use keeps the matroid structure~\cite{sonmez2022,ando2026}.
For filtered vector search, the single-category case of our constraint, Filtered-DiskANN~\cite{gollapudi2023filtered}, NHQ~\cite{wang2023nhq} and ACORN~\cite{patel2024acorn} design predicate-aware graph indexes and searches.
We take the search as given and decide when it stops, so our results apply to any search that visits items one at a time and maintains their greedy answer.

\paragraph{Early stopping in query processing and optimal stopping}
Instance-optimal aggregation stops a top-$k$ scan over sorted lists~\cite{fagin2003}, probabilistic pruning drops candidates that are unlikely to qualify~\cite{theobald2004topk}, and PAC nearest-neighbor queries stop on the distance distribution of the query~\cite{ciaccia2000pac}.
A graph search offers neither a proof that the answer is final nor a distance model, so our rule is calibrated on samples.
In optimal stopping, martingale duality bounds the best stopping rule for an unconstrained objective by simulation~\cite{rogers2002,haugh2004}, and our bound covers a risk-constrained objective over the declared class with a finite-sample guarantee.
Mixing two nested staircases is the analogue for one risk constraint.

\begin{figure*}[t]
\centering
\includegraphics[width=\textwidth]{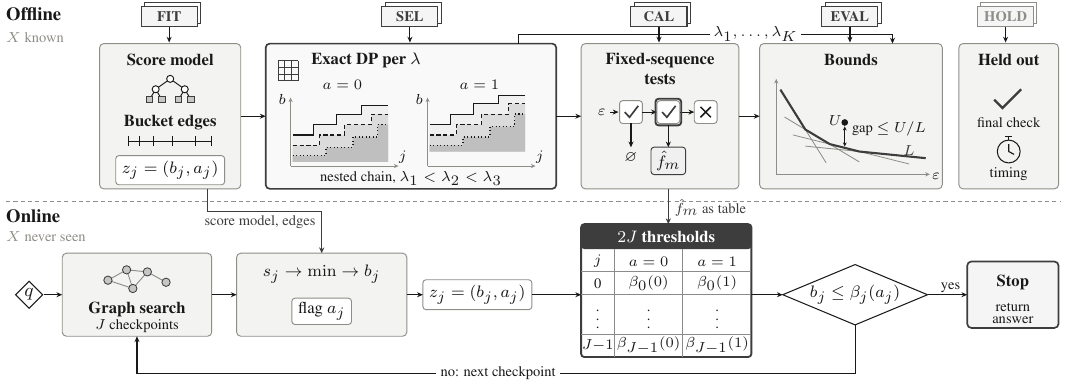}
\caption{Overview of the approach. FIT fixes the score model and the buckets of the declared interface. The monotone class yields a linear program over an order polytope, and SEL computes its nested chain of staircases. CAL selects $\hat f_m$ by fixed-sequence tests, and EVAL bounds its cost and the class optimum, which gives the ratio on the joint confidence event. Online, the search reads only the interface and a table of $2J$ thresholds.}
\label{fig:overview}
\end{figure*}

\section{Problem Formulation}\label{sec:problem}

\subsection{Constrained nearest neighbors and graph search}
Let $\mathcal V=\{1,\dots,N\}$ be a set of items.
Each item $v$ has an embedding $x_v\in\mathbb R^d$ of unit norm and a set of categories $c(v)\subseteq\mathcal C$, which may be empty.
A query $q\in\mathbb R^d$ ranks the items by the distance $d(q,x_v)=1-\langle q,x_v\rangle$, with ties broken by item index.
Capacities $\kappa:\mathcal C\to\mathbb Z_{\ge1}$ bound how many answer slots each category may fill.

\begin{definition}[Feasible answer]\label{def:feasible}
A set $S\subseteq\mathcal V$ is \emph{feasible} if there is an assignment $\phi:S\to\mathcal C$ with $\phi(v)\in c(v)$ for every $v\in S$ and $|\phi^{-1}(\gamma)|\le\kappa(\gamma)$ for every $\gamma\in\mathcal C$.
\end{definition}

Each answer item thus represents one of its categories, and no category is represented more than $\kappa$ times.
For example, with $\kappa\equiv1$, two items that both carry the categories $\{\gamma_1,\gamma_2\}$ may both be returned, one representing $\gamma_1$ and the other $\gamma_2$.
This is the semantics of a result page with one section per category, where each item appears in one section and each section shows at most $\kappa$ items.
The feasible sets are the independent sets of a transversal matroid~\cite{oxley2011}.
For $A\subseteq\mathcal V$, let $\mathrm{Greedy}(A)$ scan $A$ in increasing distance and add an item when fewer than $k$ items are kept and adding it preserves feasibility.
The \emph{exact constrained answer} is $S^\star(q)=\mathrm{Greedy}(\mathcal V)$, and we assume that $k$ does not exceed the rank of the matroid, which holds in all our settings.
The feasible sets of at most $k$ items form a matroid, so $S^\star(q)$ is a feasible set of $k$ items with the least total distance~\cite{oxley2011}, made unique by the order of the scan.
Two common queries are special cases.
Unconstrained $k$-NN gives every item one shared category with capacity $k$.
Attribute-filtered $k$-NN~\cite{gollapudi2023filtered,patel2024acorn} gives the items that pass the filter one shared category with capacity $k$ and the others none.

A graph search visits items one at a time by a best-first traversal of a proximity graph from an entry point.
Each visit computes one distance.
Let $A_t$ be the set of items visited after $t$ visits.
The search maintains $\mathrm{Greedy}(A_t)$ exactly as visits arrive, and stopping after $t$ visits returns it.
Visits arrive out of distance order, so a new item may displace a farther member of the answer.
Our engine keeps the answer with one assignment and decides each new item by an alternating-path search between answer items and category slots, and its cost is part of every reported latency.
Whether the returned answer is exact then depends only on which items have been visited, as the following lemma shows.
It lets us express the failure of a stopping rule through a single completion time.

\begin{lemma}\label{lem:visited}
If $S^\star(q)\subseteq A$, then $\mathrm{Greedy}(A)=S^\star(q)$.
\end{lemma}

\noindent Every item outside $A$ is rejected by the scan over $\mathcal V$, because every item that scan keeps belongs to $S^\star(q)\subseteq A$.
By induction over the distance order, both scans therefore hold the same kept items at every item of $A$ and make the same decision on it.
We define the \emph{completion time} $X(q)=\min\{t:S^\star(q)\subseteq A_t\}$, with $X(q)=\infty$ when no such $t$ exists.
By the lemma, stopping after $t$ visits returns the exact answer if and only if $t\ge X(q)$.
The matroid structure gives more: a visited member of the exact answer is never displaced.

\begin{lemma}\label{lem:kept}
For every $A\subseteq\mathcal V$, $S^\star(q)\cap A\subseteq\mathrm{Greedy}(A)$.
\end{lemma}

\noindent Let $s\in S^\star(q)\cap A$, and work in the matroid truncated to rank $k$.
Every item nearer than $s$ lies in the closure of the members of $S^\star(q)$ nearer than $s$, and $s$ does not.
The items that the scan over $A$ keeps before $s$ therefore cannot span $s$, so the scan keeps $s$.
Hence the recall $|\mathrm{Greedy}(A_t)\cap S^\star(q)|/k$ of the returned answer equals the fraction of $S^\star(q)$ visited, and it never decreases.
A recall target $\rho\in(0,1]$ is missed exactly when $t<X_\rho(q)$, the visit at which $\lceil\rho k\rceil$ members of $S^\star(q)$ have been visited.
Every result below thus holds with $X_\rho$ in place of $X=X_1$, and we report exactness and $\rho=0.9$.

\subsection{Checkpoints and cost}
A stopping rule acts at $J$ checkpoints, defined by a grid of visit counts $g_0<g_1<\dots<g_{J-1}$.
An \emph{expansion} removes the closest unexpanded candidate from the list and visits its unvisited neighbors.
Checkpoint $j$ is the first expansion at which the number of visits reaches $g_j$, and $c_j$ denotes the number of visits at that moment.
If the search ends before it reaches $g_j$, checkpoint $j$ and all later ones are placed at its end, with $c_j$ equal to the final number of visits.
One expansion adds at most $\Delta$ visits, where $\Delta$ is the maximum out-degree of the graph ($\Delta=64$ for our graphs).
Hence $0<c_0\le\dots\le c_{J-1}\le H=g_{J-1}+\Delta$ on every trajectory, and $H$ bounds the cost of every rule.
Stopping at checkpoint $j$ fails exactly when $c_j<X(q)$.

\subsection{Declared interface and monotone rules}
A rule observes the search only through a declared interface.
A score model, fit once on separate queries, maps the trajectory up to checkpoint $j$ to a score $s_j\in[0,1]$, where a low score indicates that the answer is likely complete.
Let $m_j=\min_{i\le j}s_i$ be its running minimum.
Fixed edges $e_1<\dots<e_{B-1}$ map $m_j$ to the bucket $b_j=|\{i:e_i\le m_j\}|\in\{0,\dots,B-1\}$, with the same edges at every checkpoint.
The flag $a_j\in\{0,1\}$ records whether the current answer holds $k$ items.
The rule observes $z_j=(b_j,a_j)\in\mathcal Z=\{0,\dots,B-1\}\times\{0,1\}$.
Along every trajectory, $b_j$ is nonincreasing in $j$ and $a_j$ is nondecreasing in $j$.
The bucket cannot rise because $m_j$ is a running minimum and the edges are fixed.
The flag cannot fall because $\mathrm{Greedy}(A_t)$ holds $\min(k,r(A_t))$ items, where $r$ is the matroid rank, which never decreases as items are visited.

\begin{definition}[Monotone stopping rules]\label{def:class}
A \emph{rule} is a sequence $f=(f_j)_{j<J}$ of maps $f_j:\mathcal Z\to[0,1]$ with $f_{J-1}\equiv1$ and
\begin{equation}\label{eq:mono}
f_j(b,a)\le f_{j'}(b',a')\quad\text{whenever } j\le j',\ b\ge b',\ a\le a'.
\end{equation}
For each query, the rule draws $\xi\sim\mathrm{Unif}[0,1]$ once, independently of the search trajectory, and stops at the first checkpoint $T$ with $f_{T}(z_{T})\ge\xi$.
We write $\Mcl$ for the set of all rules.
\end{definition}

A monotone rule stops earlier when the running minimum of the score is lower and when the answer is full, and its stopping probability does not decrease from one checkpoint to the next.
Because $z_j$ moves upward in the order of~\eqref{eq:mono} along every trajectory, $f_j(z_j)$ is nondecreasing in $j$, and $\Prob(T\le j\mid\text{trajectory})=f_j(z_j)$.
The cost and the risk of a rule are
\begin{equation}\label{eq:costrisk}
C(f)=\E\big[c_T\big],\qquad R(f)=\Prob\big(c_T<X\big),
\end{equation}
where the expectation is over the query distribution and $\xi$.
Given a risk level $\varepsilon\in[0,1]$, the problem is
\begin{equation}\label{eq:opt}
\OPT_Z(\varepsilon)=\min\{\,C(f):f\in\Mcl,\ R(f)\le\varepsilon\,\}.
\end{equation}
Waiting until the last checkpoint minimizes the risk on every trajectory, so \eqref{eq:opt} is feasible if and only if $r_0=\Prob(X>c_{J-1})\le\varepsilon$.
We set $\OPT_Z(\varepsilon)=+\infty$ otherwise.

The class $\Mcl$ contains the common practical rules once their parameters are aligned to the interface.
These include a threshold on the running minimum of the score at a bucket edge, the same threshold with a cap at a checkpoint, and a fixed budget at a checkpoint.
Every statement below is relative to the declared interface.
Rules that read other trajectory features lie outside $\Mcl$, and we compare against them in experiments.

\subsection{Samples and confidence}
We use five disjoint samples of queries, each consisting of independent draws from the query distribution.
FIT trains the score model and sets the bucket edges, SEL fits the candidate rules, CAL calibrates them, and EVAL computes the bounds.
HOLD, which no fitting or calibration step reads, serves a final check and the timing.
Probability statements are over the draws of CAL and EVAL and over the stopping draws, conditional on FIT and SEL. Refitting the score or the chain therefore requires new CAL and EVAL samples.
Each setting, a dataset searched on one graph, carries its own statements at a joint level $1-\alpha_L-\alpha_U-\delta$, and statements of different settings are not joint.
Here $\delta$, $\alpha_U$ and $\alpha_L$ are the error levels of the calibration, of the upper bound on cost and of the lower bound on the optimum. Fig.~\ref{fig:overview} summarizes the approach.

\section{The Structure of Monotone Stopping}\label{sec:structure}
Both quantities in~\eqref{eq:costrisk} are expectations at a stopping time, so their linearity in $f$ is not immediate.
The following proposition shows that they are affine.
It turns~\eqref{eq:opt} into a linear program, and every result below builds on this form.

\begin{proposition}[Affine representation]\label{prop:affine}
For every $f\in\Mcl$,
\begin{align*}
C(f)&=\E\,c_{J-1}-\textstyle\sum_{j<J-1}\E\big[f_j(z_j)\,(c_{j+1}-c_j)\big],\\
R(f)&=r_0+\textstyle\sum_{j<J-1}\E\big[f_j(z_j)\,\mathbf 1\{c_j<X\le c_{j+1}\}\big].
\end{align*}
\end{proposition}

\noindent The proof telescopes the stopping probabilities $\Prob(T\le j\mid\text{trajectory})=f_j(z_j)$.
Stopping at checkpoint $j$ instead of $j+1$ saves $c_{j+1}-c_j$ visits, and it causes a failure exactly when the answer completes between the two checkpoints.
On a sample, the empirical version of~\eqref{eq:opt} is therefore a linear program in the variables $f_j(z)$, with the monotonicity~\eqref{eq:mono} as difference constraints and one risk constraint.

\subsection{Order polytope and staircase rules}
The linear program has $(J-1)|\mathcal Z|$ free variables, and nothing so far prevents its optimum from being fractional in every coordinate.
The next theorem shows that the region of monotone rules, before the risk constraint is imposed, is an order polytope~\cite{stanley1986}, and this fixes the form of optimal rules.
It justifies the staircase form of the deployed rule and shows why exhaustive search over deterministic rules is out of reach.
Let $\preceq$ be the product order on cells $u=(j,b,a)$ in~\eqref{eq:mono}.

\begin{theorem}[Order polytope]\label{thm:polytope}
Let $P=\{x\in[0,1]^{(J-1)\times\mathcal Z}: x_u\le x_v \text{ for all } u\preceq v\}$.
\begin{enumerate}
\item[(a)] Every vertex of $P$ is a 0/1 vector. Hence every linear objective over $\Mcl$ is minimized by a deterministic rule.
\item[(b)] The deterministic rules are exactly the \emph{staircases} $f_j(b,a)=\mathbf 1\{b\le\beta_j(a)\}$ with $\beta_j(a)\in\{-1,\dots,B-1\}$, $\beta_j(a)\le\beta_{j+1}(a)$, $\beta_j(0)\le\beta_j(1)$ and $\beta_{J-1}\equiv B-1$.
\item[(c)] If~\eqref{eq:opt} is feasible, some optimum is a staircase or a mixture $(1-\theta)f^-+\theta f^+$ of two staircases with $f^-\le f^+$ in every coordinate.
\item[(d)] The number of staircases is $N_Z=\binom{B+J-1}{J-1}\binom{B+J}{J-1}/J$.
\end{enumerate}
\end{theorem}

\noindent For (b), a 0/1 point of $P$ is the indicator of an up-set, which thresholds describe. For (a), every $x\in P$ equals $\int_0^1\mathbf 1\{x\ge u\}\,du$, and every level set $\mathbf 1\{x\ge u\}$ is a 0/1 point of $P$.
For (c), the risk constraint cuts $P$ by one half-space, so every new vertex lies on an edge of $P$.
The two endpoints of an edge of an order polytope are comparable, that is, one is at most the other in every coordinate.
Indeed, if the up-sets $F$ and $G$ of two staircases are incomparable, then $\mathbf 1_F+\mathbf 1_G=\mathbf 1_{F\cup G}+\mathbf 1_{F\cap G}$, so the segment between $\mathbf 1_F$ and $\mathbf 1_G$ shares its midpoint with another segment of $P$ and is not an edge.
For (d), staircases correspond to plane partitions in a $(J-1)\times2\times B$ box, which MacMahon's box formula counts~\cite{macmahon1916}.
A staircase stops as soon as the bucket of the running minimum is at most a threshold, and the threshold grows with the checkpoint and is larger once the answer is full.
With the interface of our experiments, $J=100$ and $B=100$, so $\ln N_Z\approx266$.

\subsection{Lagrangian chain}
For $\lambda\ge0$, let $g(\lambda)=\min_{f\in\Mcl}\big[C(f)+\lambda R(f)\big]$.
A lower bound on $\OPT_Z$ must hold at every risk level a user may choose.
The lemma below reduces all risk levels to the values $g(\lambda)$, and our lower bound uses only its first part.

\begin{lemma}[Duality]\label{lem:dual}
For all $\varepsilon\in[0,1]$ and $\lambda\ge0$, $\OPT_Z(\varepsilon)\ge g(\lambda)-\lambda\varepsilon$.
If $\varepsilon\ge r_0$, equality holds for some finite $\lambda$, and $\varepsilon\mapsto\OPT_Z(\varepsilon)$ is convex, nonincreasing and piecewise linear on $[r_0,1]$.
\end{lemma}

\noindent The inequality holds because every feasible $f$ satisfies $C(f)\ge C(f)+\lambda(R(f)-\varepsilon)\ge g(\lambda)-\lambda\varepsilon$.
The equality is strong duality of the finite linear program, which needs no interior point and so also holds at $\varepsilon=r_0$.

Theorem~\ref{thm:polytope} still leaves about $e^{266}$ staircases.
The next theorem shows that the staircases relevant to~\eqref{eq:opt} form a single chain indexed by the multiplier.
This makes calibration a one-dimensional search, and it lets one dynamic program per multiplier evaluate $g(\lambda)$ exactly.
Identify a staircase with the up-set $S$ of cells $(j,b,a)$ at which it stops.
By Proposition~\ref{prop:affine}, its Lagrangian value is $\mathrm{const}+\sum_{u\in S}\theta_u(\lambda)$ with
\begin{equation}\label{eq:theta}
\begin{aligned}
\theta_{(j,b,a)}(\lambda)={}&-\E\big[(c_{j+1}-c_j)\mathbf 1_{z_j=(b,a)}\big]\\
&+\lambda\,\E\big[\mathbf 1_{c_j<X\le c_{j+1}}\mathbf 1_{z_j=(b,a)}\big]
\end{aligned}
\end{equation}
for $j<J-1$, which is nondecreasing in $\lambda$.
Stopping in cell $(j,b,a)$ saves the visits counted by the first term and adds the risk counted by the second, priced at $\lambda$.

\begin{theorem}[Nested chain]\label{thm:chain}
Let $\bar S(\lambda)$ be the union of all minimizers of the Lagrangian value over staircases.
\begin{enumerate}
\item[(a)] $\bar S(\lambda)$ is itself a minimizer.
\item[(b)] If $\lambda<\lambda'$, then $\bar S(\lambda')\subseteq\bar S(\lambda)$. The sets $\bar S(\lambda)$ form a chain along which the cost weakly decreases and the risk weakly increases as $\lambda$ decreases.
\item[(c)] Let $\bar S(\lambda{+})=\lim_{\mu\downarrow\lambda}\bar S(\mu)$, which exists because the chain is finite. For every $\varepsilon\ge r_0$ and a dual optimal multiplier $\lambda^\star$, $\OPT_Z(\varepsilon)$ is attained by $\bar S(\lambda^\star{+})$ or by a mixture of $\bar S(\lambda^\star)$ and $\bar S(\lambda^\star{+})$.
\item[(d)] Statements (a) to (c) also hold for the empirical distribution of any sample.
\end{enumerate}
\end{theorem}

\noindent Up-sets are closed under union and intersection, and the Lagrangian value is modular in $S$, so the minimizers form a lattice~\cite{topkis1978}.
Since every $\theta_u$ is nondecreasing in $\lambda$, a larger multiplier can only remove cells from the maximal minimizer.
For part (c), let $\lambda^\star>0$. Both $\bar S(\lambda^\star)$ and $\bar S(\lambda^\star{+})$ minimize the Lagrangian at $\lambda^\star$, and the risk coefficients are nonnegative, so the smaller set has the smaller risk.
The risk of $\bar S(\lambda^\star{+})$ is the right derivative of the concave function $g$ at $\lambda^\star$ and that of $\bar S(\lambda^\star)$ the left one, so dual optimality of $\lambda^\star$ makes them bracket $\varepsilon$. By Lemma~\ref{lem:dual}, the mixture that meets $\varepsilon$ with equality is then optimal.
If $\lambda^\star=0$, then $\bar S(0{+})$ is a feasible rule of minimum cost. Part (d) holds because these arguments use only expectations.
The ingredients of Theorems~\ref{thm:polytope} and~\ref{thm:chain} are classical: order polytopes~\cite{stanley1986}, lattices of minimizers~\cite{topkis1978}, maximum closure~\cite{picard1976}, whose parametric form has nested solutions~\cite{gallo1989}, and the level sets of isotonic regression~\cite{barlow1972}.
What is new is that risk-constrained stopping over a declared monotone interface has this structure.

\begin{algorithm}[t]
\caption{Exact chain element for one multiplier}\label{alg:dp}
\begin{algorithmic}[1]
\Require coefficients $\theta_{(j,b,a)}(\lambda)$ on a sample of $n$ queries, stored as the integers $2^{15}n\,\theta_{(j,b,a)}(\lambda)-1$
\State $\Pi_{j,a}(p)\gets\sum_{b<p}\theta_{(j,b,a)}$ for $p=0,\dots,B$ \Comment{stop in buckets $b<p$}
\State $D_{-1}\equiv0$
\For{$j=0,\dots,J-2$}
  \State $D_j(p_0,p_1)\gets \Pi_{j,0}(p_0)+\Pi_{j,1}(p_1)+\min_{p_0'\le p_0,\ p_1'\le p_1,\ p_0'\le p_1'}D_{j-1}(p_0',p_1')$ for $p_0\le p_1$
\EndFor
\State \Return the staircase that attains $\min_{p_0\le p_1}D_{J-2}(p_0,p_1)$ by backtracking, with $\beta_j(a)=p_{a}-1$ at layer $j$ and $\beta_{J-1}\equiv B-1$
\end{algorithmic}
\end{algorithm}

Algorithm~\ref{alg:dp} computes a chain element.
The pair $(p_0,p_1)=(\beta_j(0)+1,\beta_j(1)+1)$ encodes the staircase at checkpoint $j$, and the constraints $p_0'\le p_0$, $p_1'\le p_1$ and $p_0\le p_1$ are exactly the order constraints of Theorem~\ref{thm:polytope}(b).
Evaluating the returned staircase on $\theta_u(\lambda)$ and adding $\E c_{J-1}+\lambda r_0$ gives $g(\lambda)$.
Two-dimensional prefix minima give $O(J(B+1)^2)$ time per multiplier.
Visit counts are integers and we round every multiplier to an integer, so the coefficients $n\theta_u(\lambda)$ are integers and the program is exact at each evaluated multiplier.
An integer grid of multipliers need not recover every element of the chain or the exact SEL frontier.
To return the maximal minimizer, the program uses the integer coefficients $2^{15}n\theta_u(\lambda)-1$.
Since $2^{15}$ exceeds the number $2B(J-1)$ of free cells, this keeps the original optimum and, among its minimizers, selects the one with the most stopping cells. The interface with 200 buckets uses $2^{17}$.
On SEL, the program and a general linear-programming solver give the same value to within $10^{-4}$ visits at the two multipliers we compared in each setting.

\section{Calibration and the Optimality-Gap Bound}\label{sec:bounds}
The chain gives the shape of good rules, but it is fit on SEL and its risk there is optimistic.
This section turns the chain into a rule with a risk guarantee and bounds how far that rule can be from the class optimum.

\subsection{Calibration along the chain}
We fix a sequence of candidates on SEL and select among them on CAL.
The sequence starts with the rule that waits for the last checkpoint.
It continues with the chain elements $\bar S(\lambda)$ for a grid of multipliers in decreasing order, so each candidate is cheaper and riskier than the one before.
For candidate $k$, let $S_k$ count the CAL queries on which it fails.
The calibration accepts candidate $k$ if $\Prob\big(\mathrm{Bin}(n_{\mathrm{cal}},\varepsilon)\le S_k\big)\le\delta$, stops at the first rejection, and returns the last accepted candidate.
When the first candidate is rejected, it returns no rule.
The next theorem gives the risk half of the bound for this selection.

\begin{theorem}[Fixed-sequence calibration~\cite{angelopoulos2021ltt}]\label{thm:cal}
Let the candidate sequence be fixed before CAL is drawn. Then $\Prob\big(\text{a rule }\hat f\text{ is returned and }R(\hat f)>\varepsilon\big)\le\delta$.
\end{theorem}

\noindent Let $k^\star$ be the first candidate in the sequence whose risk exceeds $\varepsilon$.
Returning any candidate with risk above $\varepsilon$ requires accepting $k^\star$, and that event has probability at most $\delta$ because $S_{k^\star}$ is binomial with a success probability above $\varepsilon$.
The argument needs neither nested candidates nor independent tests.
The test counts the failures of each candidate at its actual stopping point, so it applies to randomized candidates as well. Each query then has one stopping draw $\xi$, independent across queries and shared by all candidates.
The nested chain makes the fixed order natural, because each candidate then stops wherever its predecessor stops and possibly earlier.

\subsection{A lower bound on the class optimum at every risk level}
The optimality-gap bound needs a denominator that bounds the population optimum from below, at every risk level a user may pick after seeing the results.
The following theorem provides it.
For a rule $f$ and a query $i$, let $W_i^\lambda(f)=\E_\xi\big[c_T+\lambda\mathbf 1\{c_T<X_i\}\big]\in[0,H+\lambda]$ be its Lagrangian value and $Q_i^\lambda(f)=\E_\xi\big[(c_T+\lambda\mathbf 1\{c_T<X_i\})^2\big]$ its second moment.
Both are affine in $f$ by the argument of Proposition~\ref{prop:affine}.
Let $\widehat W^\lambda$ and $\widehat Q^\lambda$ be their averages over the $n$ EVAL queries.

\begin{theorem}[Simultaneous lower bound]\label{thm:lower}
Let $n\ge2$. Fix the interface, multipliers $0\le\lambda_1<\dots<\lambda_K$ and levels $\alpha_k\in(0,1)$ with $\sum_k\alpha_k=\alpha_L$ before EVAL is drawn.
Let $\eta_k=\sqrt{2\ln(2/\alpha_k)/(n-1)}$, $\zeta_k=7(H+\lambda_k)\ln(2/\alpha_k)/(3(n-1))$, and let $\ell_k$ be any value with
\[
\ell_k\le\min_{f\in P}\Big[\widehat W^{\lambda_k}(f)-\eta_k\sqrt{\widehat Q^{\lambda_k}(f)-\widehat W^{\lambda_k}(f)^2}\Big]-\zeta_k .
\]
Then, with probability at least $1-\alpha_L$, simultaneously for all $\varepsilon\in[0,1]$,
\begin{equation}\label{eq:lower}
\OPT_Z(\varepsilon)\ \ge\ L(\varepsilon)=\max\Big\{0,\ \max_k\big(\ell_k-\lambda_k\varepsilon\big)\Big\}.
\end{equation}
\end{theorem}

\noindent For each $k$, fix a population minimizer $f_k^\star$ of the Lagrangian at $\lambda_k$, which is a staircase by Theorem~\ref{thm:polytope}(a).
The empirical Bernstein bound of Maurer and Pontil~\cite{maurer2009} applied to the fixed rule $f_k^\star$ bounds $g(\lambda_k)$ from below by its sample mean minus a variance term and a range term.
For a deterministic rule, the sample variance equals $\frac{n}{n-1}(\widehat Q-\widehat W^2)$, and the minimum over $P$ can only be smaller.
A union bound over $k$ and Lemma~\ref{lem:dual} at every $\varepsilon$ complete the proof.
The bound needs no uniform convergence over $\Mcl$, because only the fixed rules $f_k^\star$ enter the concentration step.
Each multiplier contributes one line $\ell_k-\lambda_k\varepsilon$, and $L$ is their upper envelope, so a single bound covers the whole frontier.

The objective inside the minimum is convex in $(\widehat W,\widehat Q)$, because $\sqrt{Q-W^2}$ is concave, so we compute $\ell_k$ by Kelley's cutting-plane method.
Every cut is an affine minorant of this objective for all values of its two parameters $C_0$ and $s>0$, by $W^2\ge2C_0W-C_0^2$ and $\sqrt v\le(v+s^2)/(2s)$. Slopes computed in floating point, with a negative coefficient of $\widehat Q$, thus give valid cuts once the intercept is recomputed exactly.
Each master value is the weak-duality value of the solver's multipliers.
One iteration suffices for each $\ell_k$ in our runs, with at most 0.07 visits between the objective and the master value at the returned point.
The reported $\ell_k$ are rounded down and every $U_m$ is rounded up.

\subsection{End-to-end optimality-gap bound}
The corollary below states the guarantee for a grid of risk levels with one calibrated rule per level.

\begin{corollary}[Optimality-gap bound along the frontier]\label{cor:ratio}
Let $E=\{\varepsilon_1,\dots,\varepsilon_M\}$, and fix $F$ candidate sequences before CAL is drawn.
For each $m$, calibrate every sequence at level $\delta/(FM)$, let $\hat f_m$ be any rule they return, selected by any procedure that may read CAL but not EVAL, and let $U_m$ be the empirical Bernstein upper bound~\cite{maurer2009} on $C(\hat f_m)$, computed at level $\alpha_U/M$ from the per-query costs on EVAL averaged over the stopping draw.
With probability at least $1-\alpha_L-\alpha_U-\delta$, for every $m$ with a returned rule and $L(\varepsilon_m)>0$,
\[
R(\hat f_m)\le\varepsilon_m\qquad\text{and}\qquad\frac{C(\hat f_m)}{\OPT_Z(\varepsilon_m)}\le\frac{U_m}{L(\varepsilon_m)} .
\]
\end{corollary}

\noindent Given FIT, SEL and CAL, each returned rule is fixed with respect to EVAL.
The corollary follows from a union bound over the $FM$ calibration events, the upper bounds and the event of Theorem~\ref{thm:lower}.
Every returned rule of every sequence has risk at most $\varepsilon_m$ on the joint event, so the selection among them needs no further correction.
The rule, the upper bound and the lower bound use disjoint data except that $U_m$ and $L$ share EVAL, which the union bound allows.
The ratio is computed after calibration, so it bounds the gap of the returned rule a posteriori at the stated level.

Every step after FIT runs offline on logged trajectories, and the deployed rule is a table of $2J$ thresholds $\beta_j(a)$.

\subsection{Size of the bound}
Corollary~\ref{cor:ratio} is data dependent, so it promises no number before the samples are seen.
Such a number would depend on the local slope of the frontier, on estimation on SEL and on the spacing of the candidates.
The next result gives a worst-case rate that no method can remove.

\begin{proposition}[Unavoidable margin]\label{prop:margin}
Fix $0<\delta<1/4$, $\varepsilon\in(0,1)$ and a positive integer $D$, and put $\Lambda=D/\varepsilon$. Consider workloads with two checkpoints, a constant interface and a last checkpoint that never fails, so that a feasible rule exists. There is $n_0(\varepsilon,\delta)$ such that, for every $n\ge n_0$ and every procedure that always returns a rule of $\Mcl$ from $n$ labeled queries with risk at most $\varepsilon$ with probability at least $1-\delta$ on each such workload, some such workload, whose frontier has left derivative $-\Lambda$ at $\varepsilon$, satisfies, with probability at least $1/2$,
\[
C(\hat f)-\OPT_Z(\varepsilon)\ \ge\ \frac{\Lambda}{8}\sqrt{\frac{\varepsilon(1-\varepsilon)\log(1/(4\delta))}{n}} .
\]
\end{proposition}

\noindent\emph{Proof sketch.} A rule is an early-stop probability $t$, with risk $pt$ and cost $a+D(1-t)$ when a query completes between the two checkpoints with probability $p$. Let $d$ be one quarter of the root above. Safety at $p=\varepsilon+d$ gives $\hat t\le\varepsilon/(\varepsilon+d)$ with probability at least $1-\delta$. Since $n\,\mathrm{kl}(\varepsilon,\varepsilon+d)\le\frac18\log\frac1{4\delta}<\mathrm{kl}(\frac12,1-\delta)$, data processing keeps this event at probability at least $1/2$ under $p=\varepsilon$, where it costs at least $Dd/(2\varepsilon)$ more than the optimum.
In the local regime with fixed $M$ and a chance proportional to $\delta$ that the rule fails the test, the calibration margin has the same leading dependence on $n$, the local variance and small $\delta$. Its constant is about ten times larger. At $\varepsilon=0.05$, $n=40{,}000$ and the level $\delta/(2M)=0.05/18$, the acceptance cutoff lies about 0.003 below $\varepsilon$, whereas the bound divided by $\Lambda$ is about 0.0003.
The next lemma bounds what the spacing of the multipliers costs.

\begin{lemma}[Multiplier interval]\label{lem:interval}
Let $f^-$ and $f^+$ minimize an empirical Lagrangian at $\lambda_{\rm hi}\ge\lambda_{\rm lo}\ge0$, with empirical risks $r_-<r_+$. A mixture with empirical risk $r=r_-+\theta(r_+-r_-)$ costs at most $(\lambda_{\rm hi}-\lambda_{\rm lo})\theta(1-\theta)(r_+-r_-)$ more than the empirical optimum $\widehat{\OPT}_Z(r)$ at risk $r$.
\end{lemma}

\noindent\emph{Proof.} Optimality at the two multipliers gives $\widehat C(f^-)\le\widehat{\OPT}_Z(r)+\lambda_{\rm hi}(r-r_-)$ and $\widehat C(f^+)\le\widehat{\OPT}_Z(r)+\lambda_{\rm lo}(r-r_+)$. Weighting them by $1-\theta$ and $\theta$ and adding gives the claim, since the cost is affine in the mixture.
The whole gap need not lie within a constant of the best possible, because a deterministic chain can stay coarse even with infinite data.

\section{Experiments}\label{sec:exp}
The experiments ask how close CaST is to the best rule of its class and whether richer policies outside the class are cheaper.
They also ask whether the saving holds in time, on held-out queries, on other workloads and indexes, under a query shift and after index updates.

\subsection{Experiments Setup}
\subsubsection{Datasets}
We use four public corpora (Table~\ref{tab:data}).
\emph{Amazon} is the Sports and Outdoors category of Amazon Reviews 2023~\cite{hou2026}, where the categories of an item are the second-level node and the leaf of its category path.
\emph{arXiv} is a metadata snapshot,\footnote{\url{https://huggingface.co/datasets/librarian-bots/arxiv-metadata-snapshot}} where the categories are the subject classes of a paper.
Both are embedded with all-MiniLM-L6-v2~\cite{reimers2019}.\footnote{\url{https://huggingface.co/sentence-transformers/all-MiniLM-L6-v2}}
\emph{YFCC-10M} is the filtered-search dataset of the NeurIPS'23 Big-ANN competition~\cite{simhadri2025bigann}, with CLIP~\cite{radford2021clip} embeddings of YFCC100M images~\cite{thomee2016yfcc}.
Its categories are the 1{,}000 most frequent tags, and we also use its 200{,}000 official queries without their tag predicates.
\emph{YouTube-8M}~\cite{abuelhaija2016yt8m} holds video features, with machine-generated entity labels as categories.

The main workload asks for $k=10$ items with capacity $\kappa\equiv1$ on Amazon and arXiv.
There the constrained answer differs from the unconstrained top 10 for 98.6\% and 98.2\% of queries, against about 70\% on YouTube-8M and 4.4\% on YFCC-10M.
For Amazon and arXiv, one random permutation of the pool selects the base, 2{,}500 FIT queries and 200{,}000 further queries, which are split at random into SEL, CAL, EVAL and HOLD.
We model the pool as independent draws from an item distribution, the query distribution of our guarantees, so the base and the five samples are independent samples of it.
Exact constrained answers are computed by a full scan.
Between 0.10\% and 0.58\% of EVAL queries of the main workload do not complete within 32{,}768 visits.

Thirteen further settings test the breadth of the method.
On graph A of Amazon and arXiv, they raise the capacity to $\kappa\equiv2$, drop the constraint, which gives plain $k$-NN, or query with only the titles of the 200{,}000 further items, split again at random.
Two run the main workload on a Vamana graph of each dataset.
On YFCC-10M and YouTube-8M, they ask constrained and unconstrained queries with $k=10$, and YFCC-10M also constrained official queries.
Each workload has its own exact answers, logs and samples of the same sizes, without HOLD for the title settings, YFCC-10M and YouTube-8M.

\begin{table}[t]
\centering
\caption{Dataset statistics. Each dataset has its own queries, with the same sample sizes: FIT 2{,}500, SEL 20{,}000, CAL 40{,}000, EVAL 120{,}000 and, for Amazon and arXiv, HOLD 20{,}000.}
\label{tab:data}
\setlength{\tabcolsep}{4pt}
\begin{tabular}{@{}lrrrrr@{}}
\toprule
 & & & & \multicolumn{2}{c}{Categories per item}\\
\cmidrule(l){5-6}
Dataset & Base items & Dim. & Categories & Max & Mean\\
\midrule
Amazon & 1{,}000{,}000 & 384 & 2{,}648 & 2 & 1.98\\
arXiv  & 1{,}000{,}000 & 384 & 176 & 10 & 1.74\\
YFCC-10M & 9{,}791{,}295 & 192 & 1{,}000 & 296 & 5.54\\
YouTube-8M & 1{,}000{,}000 & 1{,}024 & 3{,}862 & 21 & 3.01\\
\bottomrule
\end{tabular}
\end{table}

\begin{table*}[t]
\centering
\caption{Visits and optimality-gap bound on EVAL, on graph A unless a row names another graph. Fewest visits per row in bold, second fewest underlined. The first four rows are the main workload, timed on HOLD for graph A. Below them, every ratio above one is significant at level $0.025/102$. $U/L$ is rounded up, its maximum over the levels from 0.03 to 0.10, on YFCC-10M from 0.05.}
\label{tab:breadth}
\footnotesize
\setlength{\tabcolsep}{4pt}
\begin{tabular}{@{}lccccccr@{\hspace{16pt}}cc@{}}
\toprule
 & \multicolumn{6}{c}{Ratio to CaST (as 1.000) at $\varepsilon=\delta=0.05$ $\downarrow$} & & \multicolumn{2}{c}{Frontier rule}\\
\cmidrule(r{8pt}){2-7}\cmidrule(l){9-10}
Workload & Fixed & LAET-style & Level & Level+cap & List size~\cite{malkov2020} & Dist.-adaptive~\cite{aljazzazi2025adaptive} & \textbf{CaST visits}$\downarrow$ & $U/L$ max$\downarrow$ & $U/L$ at 0.05$\downarrow$\\
\midrule
\multicolumn{10}{@{}l}{\emph{Exact answer}}\\
Amazon, A & 1.634 & 1.381 & 1.184 & \underline{\smash[b]{1.076}} & 1.751 & 1.139 & \textbf{4{,}155} & 1.112 & 1.087\\
Amazon, B & 1.596 & 1.340 & 1.169 & \underline{\smash[b]{1.075}} & 1.710 & 1.112 & \textbf{4{,}255} & 1.128 & 1.075\\
arXiv, A & 1.576 & 1.326 & 1.161 & \underline{\smash[b]{1.113}} & 1.507 & 1.410 & \textbf{3{,}798} & 1.082 & 1.064\\
arXiv, B & 1.551 & 1.306 & 1.097 & \underline{\smash[b]{1.037}} & 1.482 & 1.387 & \textbf{3{,}861} & 1.098 & 1.083\\
\addlinespace[2.5pt]
Amazon, $\kappa\equiv2$ & 1.617 & 1.339 & 1.267 & \underline{\smash[b]{1.080}} & 1.779 & 1.151 & \textbf{2{,}239} & 1.181 & 1.105\\
Amazon, unconstrained & 1.900 & 1.471 & 1.563 & 1.103 & 1.578 & \textbf{0.973} & \underline{\smash[b]{897}} & 1.154 & 1.146\\
arXiv, $\kappa\equiv2$ & 1.476 & 1.311 & 1.110 & \underline{\smash[b]{1.027}} & 1.482 & 1.220 & \textbf{2{,}164} & 1.083 & 1.071\\
arXiv, unconstrained & 1.598 & 1.258 & 1.098 & 1.024 & 1.410 & \textbf{0.998} & \underline{\smash[b]{1{,}461}} & 1.106 & 1.064\\
\addlinespace[2.5pt]
Amazon, Vamana & 1.452 & 1.250 & 1.350 & 1.262 & 1.567 & \underline{\smash[b]{1.123}} & \textbf{4{,}394} & 1.098 & 1.070\\
arXiv, Vamana & 1.518 & 1.257 & 1.074 & \underline{\smash[b]{1.039}} & 1.578 & 1.404 & \textbf{3{,}707} & 1.110 & 1.084\\
\addlinespace[2.5pt]
Amazon, title queries & 1.639 & 1.337 & 1.213 & \underline{\smash[b]{1.089}} & 1.689 & 1.137 & \textbf{5{,}330} & 1.111 & 1.081\\
arXiv, title queries & 1.623 & 1.279 & 1.118 & \underline{\smash[b]{1.093}} & 1.453 & 1.151 & \textbf{7{,}370} & 1.114 & 1.094\\
\addlinespace[2.5pt]
YFCC-10M, tags & 1.279 & \underline{\smash[b]{1.082}} & 1.115 & 1.086 & 1.141 & 1.166 & \textbf{16{,}306} & 1.372 & 1.372\\
YFCC-10M, unconstrained & 1.288 & 1.073 & 1.133 & 1.123 & \underline{\smash[b]{1.051}} & 1.214 & \textbf{16{,}198} & 1.364 & 1.364\\
YFCC-10M, official vectors & 1.313 & 1.090 & 1.132 & 1.132 & \underline{\smash[b]{1.016}} & 1.133 & \textbf{16{,}034} & 1.177 & 1.177\\
YouTube-8M, entities & 1.799 & 1.295 & 1.294 & \underline{\smash[b]{1.144}} & 1.388 & 1.190 & \textbf{3{,}126} & 1.103 & 1.083\\
YouTube-8M, unconstrained & 1.893 & 1.335 & 1.261 & \underline{\smash[b]{1.133}} & 1.457 & 1.136 & \textbf{2{,}309} & 1.150 & 1.150\\
\midrule
\multicolumn{10}{@{}l}{\emph{Recall target $\rho=0.9$, main workload}}\\
Amazon, A & 1.554 & 1.249 & 1.059 & \underline{\smash[b]{1.018}} & 1.725 & 1.395 & \textbf{1{,}929} & 1.105 & 1.081\\
Amazon, B & 1.580 & 1.245 & 1.080 & \underline{\smash[b]{1.030}} & 1.754 & 1.419 & \textbf{1{,}897} & 1.093 & 1.068\\
arXiv, A & 1.565 & 1.184 & 1.057 & \underline{\smash[b]{1.038}} & 1.535 & 1.748 & \textbf{1{,}800} & 1.106 & 1.058\\
arXiv, B & 1.584 & 1.204 & 1.050 & \underline{\smash[b]{1.046}} & 1.552 & 1.767 & \textbf{1{,}779} & 1.075 & 1.055\\
\bottomrule
\end{tabular}
\end{table*}

\subsubsection{Metrics}
Cost is the mean number of bottom-layer visits. The upper-layer descent adds about 170 distances per query to every rule, and latency includes it.
Risk is the fraction of queries whose returned answer differs from the exact constrained answer, or, for a recall target $\rho$, holds fewer than $\lceil\rho k\rceil$ of its members.
Latency is the mean over queries of the median wall-clock time of five passes, including the evaluation of features and models.
The optimality-gap bound is the ratio $U_m/L(\varepsilon_m)$, and 1 means optimal within the class.

\subsubsection{Compared methods}
The compared rules fall into four groups, all calibrated by the same test on CAL.

\key{The first group} reads the same seven features on the same FIT queries, which isolates the value of the staircase shape.
\emph{Level} stops when the running minimum of the score falls below a threshold, the stopping principle of DARTH applied to our score, and \emph{Level+cap} adds a visit cap chosen on SEL.
Both read the score at a finer resolution than $z_j$.
\emph{LAET-style} regresses the logarithm of the completion time at a checkpoint chosen on SEL, as LAET does, and stops at a multiple of the prediction.
\emph{Fixed} uses one visit budget.

\key{The second group} ports \emph{DARTH} and \emph{DARTH+}~\cite{chatzakis2026darthplus} into our engine, with recall measured against the exact constrained answer.
DARTH uses the authors' eleven answer features, regressor settings and adaptive intervals, trained on 10{,}000 SEL queries.
Its recall target is calibrated to 0.9942 on Amazon and 0.9939 on arXiv, above every target its authors evaluate. At the literal targets 0.90 and 0.95, the port reaches mean recalls of 0.908 and 0.956 to 0.959, the average-recall goal DARTH is designed for.
DARTH+ adds eight query statistics and stops on a learned lower quantile of the recall, with its level chosen on unused SEL queries.
Its code is not public, so we port it from the paper. We check only that the engine evaluates the fitted forests as scikit-learn does.

\key{The third group} tunes the search.
\emph{List size} runs the traversal with a list of $\mathit{ef}$ entries, replayed from the logged traversal as the top-$\mathit{ef}$ prefix of the longer list.
On the 160{,}000 CAL and EVAL queries of each main setting, native runs at the selected list size differ from the replay in the visits of at most two queries and the exactness of two.
Post-filtering the final list of the same traversal, the usual two-stage approach, is exact only when List size is, by Lemma~\ref{lem:visited}.
\emph{Distance-adaptive} adapts Adaptive Beam Search~\cite{aljazzazi2025adaptive}: it stops once the answer is full and the next candidate is farther than $1+\gamma$ times the $k$-th distance of the answer.
They are calibrated over 46 list sizes from 8{,}192 down to 10 and over 129 values of $\gamma$ from 2 down to 0, 246 on YFCC-10M.
\emph{Ada-ef}~\cite{zhang2026adaef} looks up a list size from a Gaussian model of the distances of the query, with success meaning the exact answer.
\emph{Patience}~\cite{teofili2025patience} stops once the answer is full and unchanged for a calibrated number of visits.

\key{The fourth group} holds richer learned policies, which test whether the restriction to $\Mcl$ costs much.
By Longstaff-Schwartz style backward induction~\cite{longstaff2001}, gradient-boosted models estimate the further cost of continuing and the chance that it turns a failure into an exact answer. A policy stops when the former exceeds $\lambda'$ times the latter.
The fully fitted policy reads the seven features, the score and its running minimum on all of SEL. The cross-fitted policy fits each half of SEL on the other half, and the declared-state policy reads only $z_j$.

\subsubsection{Implementation details}
\key{Index and interface.} Amazon and arXiv are each indexed by two HNSW graphs~\cite{malkov2020} (degree parameter 32, construction list 200), built with two independent random draws of the insertion levels.
We call them graphs A and B, and a dataset with a graph a \emph{setting}. YFCC-10M and YouTube-8M have one such graph each.
Amazon and arXiv also have a Vamana graph built with the DiskANN reference implementation~\cite{subramanya2019}, with out-degree 64, construction list 200 and $\alpha=1.2$.
The search is a best-first traversal with a list of 8{,}192 entries that maintains the greedy constrained answer of the visited items.
As in HNSW, it ends when the closest unexpanded candidate is farther than all 8{,}192 kept entries.
The rule acts at $J=100$ checkpoints on a geometric grid from 64 to 32{,}768 visits, so $H=32{,}832$.
The score model is a gradient-boosted classifier (at most 200 iterations, 31 leaves), trained on FIT to predict whether the answer is already complete, and the score is one minus its predicted probability.
Its seven features at a checkpoint describe the visits, the distance of the candidate about to be expanded and the size of the answer. They add the $k$-th distances of the answer and of the unconstrained top $k$, and the logarithm and the fraction of visits since the answer changed.
Its running minimum is bucketed at the 99 percentiles of the FIT scores pooled over all checkpoints, so $B=100$.

\key{Calibration.} Our rule is calibrated in two ways.
For the frontier results, we apply Corollary~\ref{cor:ratio} with $M=9$ risk levels, 0.01 to 0.08 and 0.10, $\delta=0.05$ and $F=2$ candidate sequences. Each calibration thus runs at level $\delta/(2M)$ and returns the \emph{frontier rule} of its level.
The two sequences are the chain and the mixture sequence, optimal solutions of program~\eqref{eq:opt} on SEL at 40 fixed multiples of $\varepsilon$, both fixed before CAL is read, and we report the chain.
The levels and the reporting range from 0.03 to 0.10 were set before these calibrations, and development read CAL only to compare the two sequences, which $F=2$ covers. A bound at 0.01 from an earlier version of the method, on the same EVAL, was known and partly motivated this floor, so we report 0.01 and 0.02 as well.
The bounds use $\alpha_U=0.025$, split evenly over the $M$ cost bounds, and $\alpha_L=0.05$, split evenly over $K=9$ multipliers, so each setting carries a joint level of 87.5\%. The multipliers are the negated slopes of the SEL frontier at the nine levels.
The ratios of the four main settings and of Table~\ref{tab:breadth} use exact rational arithmetic. Ratios at other confidence levels, samples or interfaces use floating point with a relative margin of $10^{-6}$.
For the comparison with the other rules, every rule, ours included, is calibrated by the same fixed-sequence test on CAL at $\varepsilon=0.05$ and $\delta=0.05$.
Unless stated otherwise, comparisons and timings use this rule, and every bound $U/L$ is that of the frontier rule.
Offline, the dynamic programs on SEL, the calibration on CAL and the bounds on EVAL take 12 to 13 minutes per setting on one server.

\key{Timing.} Latency is measured on one pinned core of a 16-core desktop processor, with the passes in rotated order.
A timing window is one such session. Ratios are formed within a window only, because the staircase moves by up to 3.9\% between the first three windows and by 9\% in the fourth.
The list size is calibrated to $\mathit{ef}=662$ and 319 on Amazon and arXiv, and to 627 and 314 on a finer grid in the third window.
\emph{Significantly slower} means that a one-sided paired bootstrap test at the 2.5\% level rejects a ratio of at most one.
In the third and fourth windows, every such result holds after a Bonferroni correction within the window.
The level $0.025/102$ of Table~\ref{tab:breadth} uses a normal approximation with the bootstrap standard error.

\key{Deployed rule.} Calibration and the bounds use outcomes computed offline from logged trajectories.
On 10{,}000 EVAL queries of Amazon and arXiv, the online stopping point of each rule checked equals the offline one on all but at most five queries, and exactness always agrees.
For our rule at $\varepsilon=0.05$ on graph A, under either calibration, the stopping point differs on 1 of the 30{,}000 EVAL and HOLD queries of Amazon and on none of arXiv.
An exact one-sided binomial bound at level 0.05 then limits the probability $p$ that the deployed rule and its offline replay differ to 0.016\%.
The deployed rule therefore has risk at most $\varepsilon+p$ and cost at most $U_m+Hp$. At 0.05 this raises the ratio by at most 0.004, at a joint level lowered by 0.05.
Our public repository (\url{https://github.com/zhangshea/CaST}) holds the code and documents all implementation details, including how the random draws were made.

\begin{table}[t]
\centering
\caption{Mean latency, visits and failure rate (\%) on the 20{,}000 HOLD queries of graph A, every rule calibrated at $\varepsilon=\delta=0.05$. Ratio is a rule's latency over ours within a timing window. Best in bold, second underlined. Fail, not ranked, stays below 5.26, where an exact binomial test at level 0.05 rejects the target.}
\label{tab:latency}
\footnotesize
\setlength{\tabcolsep}{2pt}
\begin{tabular}{@{}lrrrr@{\hspace{7pt}}rrrr@{}}
\toprule
 & \multicolumn{4}{c}{Amazon} & \multicolumn{4}{c}{arXiv}\\
\cmidrule(lr){2-5}\cmidrule(l){6-9}
Rule & $\mu$s$\downarrow$ & Visits$\downarrow$ & Fail & Ratio$\downarrow$ & $\mu$s$\downarrow$ & Visits$\downarrow$ & Fail & Ratio$\downarrow$\\
\midrule
Fixed & 1{,}715 & 6{,}789 & 4.74 & 1.440 & 1{,}446 & 5{,}987 & 4.49 & 1.402\\
LAET-style~\cite{li2020laet} & 1{,}532 & 5{,}735 & 5.02 & 1.286 & 1{,}298 & 5{,}039 & 4.87 & 1.258\\
Level & 1{,}398 & 4{,}930 & 5.04 & 1.174 & 1{,}192 & 4{,}370 & 4.91 & 1.155\\
Level+cap & \underline{\smash[b]{1{,}285}} & \underline{\smash[b]{4{,}475}} & 5.19 & \underline{\smash[b]{1.079}} & \underline{\smash[b]{1{,}127}} & \underline{\smash[b]{4{,}210}} & 4.82 & \underline{\smash[b]{1.092}}\\
DARTH~\cite{chatzakis2025darth} & 1{,}725 & 6{,}693 & 5.10 & 1.448 & 1{,}292 & 5{,}106 & 4.83 & 1.252\\
\midrule
\textbf{CaST (ours)} & \textbf{1{,}191} & \textbf{4{,}166} & 5.07 & \textbf{1.000} & \textbf{1{,}032} & \textbf{3{,}780} & 4.96 & \textbf{1.000}\\
\midrule
\multicolumn{9}{@{}l}{\emph{Second timing window}}\\
List size~\cite{malkov2020} & 1{,}740 & 7{,}289 & 4.78 & 1.413 & 1{,}273 & 5{,}715 & 4.51 & 1.239\\
Level+cap & \underline{\smash[b]{1{,}336}} & \underline{\smash[b]{4{,}475}} & 5.19 & \underline{\smash[b]{1.085}} & \underline{\smash[b]{1{,}124}} & \underline{\smash[b]{4{,}210}} & 4.82 & \underline{\smash[b]{1.093}}\\
\textbf{CaST (ours)} & \textbf{1{,}231} & \textbf{4{,}166} & 5.07 & \textbf{1.000} & \textbf{1{,}028} & \textbf{3{,}780} & 4.96 & \textbf{1.000}\\
\midrule
\multicolumn{9}{@{}l}{\emph{Third timing window}}\\
DARTH~\cite{chatzakis2025darth} & 1{,}725 & 6{,}693 & 5.10 & 1.455 & 1{,}265 & 5{,}106 & 4.83 & 1.254\\
DARTH+~\cite{chatzakis2026darthplus} & 1{,}658 & 6{,}331 & 4.86 & 1.398 & 1{,}270 & 5{,}088 & 4.72 & 1.260\\
List size~\cite{malkov2020} & 1{,}620 & 6{,}969 & 5.06 & 1.366 & 1{,}264 & 5{,}644 & 4.59 & 1.254\\
Patience~\cite{teofili2025patience} & 1{,}480 & 5{,}826 & 4.54 & 1.249 & 1{,}222 & 5{,}108 & 4.23 & 1.212\\
Dist.-adaptive~\cite{aljazzazi2025adaptive} & \underline{\smash[b]{1{,}243}} & 4{,}728 & 5.10 & \underline{\smash[b]{1.048}} & 1{,}312 & 5{,}289 & 4.89 & 1.301\\
Level+cap & 1{,}283 & \underline{\smash[b]{4{,}475}} & 5.19 & 1.082 & \underline{\smash[b]{1{,}103}} & \underline{\smash[b]{4{,}210}} & 4.82 & \underline{\smash[b]{1.094}}\\
\textbf{CaST (ours)} & \textbf{1{,}186} & \textbf{4{,}166} & 5.07 & \textbf{1.000} & \textbf{1{,}008} & \textbf{3{,}780} & 4.96 & \textbf{1.000}\\
\midrule
\multicolumn{9}{@{}l}{\emph{Fourth timing window, beside other processes}}\\
Curve (in class) & \underline{\smash[b]{1{,}275}} & \underline{\smash[b]{4{,}410}} & 4.83 & \underline{\smash[b]{1.058}} & \underline{\smash[b]{1{,}166}} & \underline{\smash[b]{4{,}025}} & 5.00 & \underline{\smash[b]{1.061}}\\
Level+cap & 1{,}321 & 4{,}475 & 5.19 & 1.096 & 1{,}185 & 4{,}210 & 4.82 & 1.078\\
\textbf{CaST (ours)} & \textbf{1{,}205} & \textbf{4{,}166} & 5.07 & \textbf{1.000} & \textbf{1{,}099} & \textbf{3{,}780} & 4.96 & \textbf{1.000}\\
\bottomrule
\end{tabular}
\end{table}

\subsection{Experimental Results and Analysis}
\subsubsection{Overall comparison}
\begin{figure}[t]
\centering
\includegraphics[width=\columnwidth]{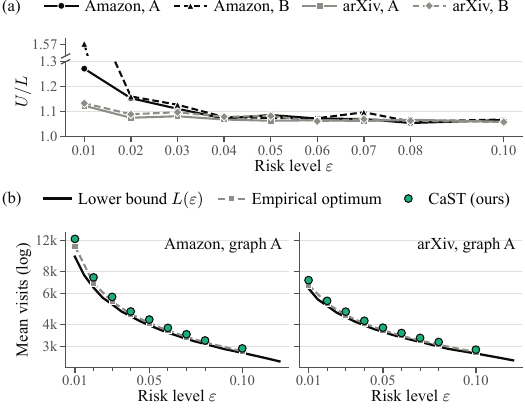}
\caption{Near-optimality on EVAL of the frontier rule, calibrated at level $\delta/(2M)$. (a)~Optimality-gap bound $U/L$ (lower is better, and 1 is optimal within the class), rounded up and holding jointly over the nine levels at 87.5\% per setting. (b)~Cost-risk frontier of graph A. The lower bound $L(\varepsilon)$ holds for all risk levels at once, the empirical class optimum is solved exactly on EVAL, and each marker is the mean visits of our rule at one risk level.}
\label{fig:nearopt}
\end{figure}

\paragraph{Near-optimality}
\key{CaST stays close to the best rule of its class.}
From 0.03 to 0.10, the maximum bound $U/L$ of each setting ranges from 1.082 to 1.128 at joint level 87.5\% (Fig.~\ref{fig:nearopt}(a)), so CaST costs at most 1.128 times the class optimum.
At 95\%, with $\delta=0.02$, $\alpha_U=0.01$ and $\alpha_L=0.02$, these maxima rise by at most 0.004, to 1.132, and the rule calibrated at 0.05 stays the same.
These bounds are for the rule replayed on the logged searches.
The failure rate on EVAL stays at 75\% to 97\% of the target in every setting.
On the frontier of graph A in Fig.~\ref{fig:nearopt}(b), the rule stays close to both the empirical optimum and the lower bound from 0.03 upward.
At 0.02 the ratio is 1.076 to 1.161 across the four settings, and at 0.01 it is 1.124 to 1.572, the largest on Amazon graph B.

\key{Over other samples, the maximum at 87.5\% stays at most 1.223.}
With three further disjoint FIT samples of 2{,}500 queries on graph A, each defining its own interface and class, it is 1.085 to 1.142.
Over ten further random splits of SEL, CAL and EVAL on graph A, with FIT and HOLD fixed, it is 1.094 to 1.223 on Amazon and 1.070 to 1.165 on arXiv.
Its medians are 1.145 and 1.100, against 1.112 and 1.082 on the reported split, which thus lies on the favorable side.
In every split the failure rate stays below the target at all nine levels.

\key{Little remains to be gained within the class.}
On the joint event, no rule of the class with risk at most 0.05 saves more than a fraction $1-L/U$ of the cost of our rule, 6.0\% to 8.0\% by setting.
Over the 28 pairs of setting and level from 0.03 to 0.10, the cost bound, the lower bound and the displacement of the risk budget take 72.5\% to 94.2\% of the gap $U-L$. The rule against the empirical frontier takes 4.7\% to 24.2\% and the multiplier grid at most 4.5\%.
At 0.01 on Amazon graph B, the mixture sequence lowers the ratio to 1.32 and moves the other three settings by at most 0.03.

Outside the class, a clairvoyant rule stops at the first checkpoint after completion and spends the risk budget on the queries of largest avoidable cost.
At 0.05 it is 2.8 to 3.3 times cheaper than the empirical class optimum, which caps what any richer interface could save on the same checkpoints.
Against the clairvoyant rule, $U/L$ bounds the loss of our rule within the interface on the joint event, and this in-sample factor estimates the loss of the interface. When the first is near one, only a rule outside the class can help, through new observations or a non-monotone use of the same ones.

\paragraph{Workload breadth}
\key{The near-optimality and the saving in visits carry over to most further settings.}
Outside YFCC-10M, the ratio in Table~\ref{tab:breadth} stays at most 1.181 at every level from 0.03 to 0.10, and at most 1.110 on the two Vamana graphs.
Title queries need 1.28 and 1.94 times the visits of item queries at 0.05, yet refit on them the ratio stays at most 1.114.
At $\varepsilon=0.05$ our rule is the cheapest in eleven of the thirteen settings.
In the other two, unconstrained Amazon and arXiv, the distance-adaptive rule, which was designed for plain $k$-NN, is 2.7\% and 0.2\% cheaper at a similar failure rate.

On YFCC-10M, 3.9\% of the queries do not complete within 32{,}768 visits, so we evaluate the levels from 0.05 to 0.10 only.
There the ratio is 1.364 and 1.372 at 0.05, just above this share, and it falls to 1.124 and 1.125 at 0.10.
On 10{,}000 EVAL queries of YFCC-10M with tags, the first group takes 1.2\% to 22.7\% more time than our rule.
The list size takes 12.1\% less, at 0.18\,$\mu$s per visit against our 0.24\,$\mu$s on the list of 8{,}192.
With a list of 2{,}048, chosen among four sizes without reading CAL and with both bounds split over the four, our staircase is within 1\% of the list size in time. Its bound is 1.402 at 0.05 and 1.129 at 0.10.
On the main workload, the same change makes our rule 1.2\% and 2.2\% slower on HOLD. With $k=50$, 15.8\% and 30.2\% of the queries of Amazon and arXiv do not complete within the horizon, above every evaluated risk level.

\paragraph{Recall target}
\key{Under the recall target $\rho=0.9$, our rule is again the cheapest, and every other rule of Table~\ref{tab:breadth} is significantly costlier at $\varepsilon=0.05$.}
By Lemma~\ref{lem:kept}, the pipeline applies with the completion time $X_{0.9}$ and a score model retrained on the new label.
On the main workload, the ratio stays at most 1.106 at every level from 0.03 to 0.10, and at 0.05 the staircase needs less than half of its exact-answer visits.
On graph A, DARTH and DARTH+, which predict recall, spend 1.23 to 1.60 times our visits under the same test.

\begin{figure}[t]
\centering
\includegraphics[width=\columnwidth]{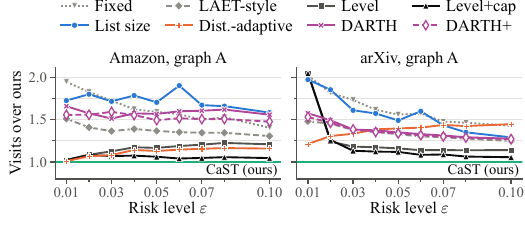}
\caption{Visits of each compared rule over ours on the 120{,}000 EVAL queries of graph A of the main workload, every rule calibrated at each risk level at $\delta/(2M)$ like the frontier rule, not at $\delta=0.05$ as in Table~\ref{tab:breadth}.}
\label{fig:levels}
\end{figure}

\paragraph{Cost and latency at a common risk target}
\key{At $\varepsilon=0.05$, our rule has the lowest mean latency in every timing window.}
On the 20{,}000 HOLD queries, every other calibrated rule takes significantly more mean time than ours, by 4.8\% to 45.5\% (Table~\ref{tab:latency}).
The saving in visits carries over to end-to-end time, and the stopping check of the staircase, score model included, takes 7.5\% and 7.1\% of its query time.
At 16 threads, our throughput over that of each rule of the third window is 1.07 to 1.66, within 0.06 of the visit ratio, so the saving holds under load.
These savings are in the mean. The distance-adaptive rule and, on Amazon, Level+cap have a lower median. In the third window, our 99th percentile is 3{,}423 and 3{,}417\,$\mu$s, above the list size (2{,}670 and 1{,}811), Patience (3{,}033 and 2{,}682) and, on arXiv, Level+cap (2{,}449).

\key{In visits, our rule is the cheapest at every level on graph A (Fig.~\ref{fig:levels}).} On graph B, the six rules of Table~\ref{tab:breadth} calibrated the same way spend at least 2.5\% more visits at every level from 0.02 upward, so the advantage does not hinge on the chosen risk level.
Cheaper rules appear only at 0.01 on Amazon graph B, where the distance-adaptive rule spends 0.80 times our visits and Level and Level+cap 0.84 times.
On graph A, the gap to Fixed is largest at 0.01. One budget for all queries must reach the $(1-\varepsilon)$-quantile of the completion time, while the staircase stops each query by its own progress.

\key{The saving does not come from failing more often.} Interpolated to equal EVAL failure rates from 0.03 to 0.08, the first group still spends 1.04 to 1.68 times our visits and the distance-adaptive rule 1.08 to 1.44 times. Patience, swept over its parameter, spends 1.17 to 1.40 times, and DARTH and DARTH+ on graph A 1.28 to 1.63 times.
On all EVAL queries, Patience and DARTH+ spend 1.32 to 1.52 times our visits wherever we run them, each significantly more.
Ada-ef passes the test only with its table built on all of SEL, at 1.49 to 1.81 times our visits.

\key{The staircase shape explains part of our saving over Level+cap.} At $\varepsilon=0.05$ on Amazon, the calibrated staircase stops only in buckets up to 20 until 702 visits and then relaxes its threshold step by step. On arXiv it never stops before 1{,}318 visits.
Level keeps one threshold for the whole search and Level+cap adds a single cap, so neither can follow such a profile.
A threshold curve of the class with an offset, a break point, a slope, a flag gap and a cap can follow it.
Fit on SEL and calibrated by the same test, it spends 0.976 to 1.074 times our visits over ten random splits per setting, at failure rates within 0.3 points.
Its median per setting is 1.012 to 1.053, and it recovers a median of 41\% to 85\% of the gap to Level+cap.
The rest of this gap comes from long queries, where the thresholds of the chain rise faster than the single slope of the curve.
On HOLD, the curve of the reported split takes 5.8\% and 6.1\% more mean time than ours, within 0.004 of its visit ratio.
The reported split favors our rule only on Amazon graph A, where its EVAL visit ratio of 1.059 exceeds all ten.

\paragraph{Held-out check}
\key{On the held-out queries, no test rejects the risk target of the frontier rule in any setting or at any level.}
In each of the four settings, the rule at $\varepsilon=0.05$, calibrated at level $\delta/(2M)$, fails on at most 995 of the 20{,}000 HOLD queries. An exact binomial test at level $0.05/4$ would need 1{,}071 failures to reject the target.
At all nine levels, fixed before HOLD was read, a joint test of the 36 pairs of setting and level at level $0.05/36$ each rejects none. Recomputed on HOLD alone, a sixth the size of EVAL, the bound is at most 1.217 from 0.03 to 0.10 and 1.897 at 0.01.
In every setting, the Maurer-Pontil lower bound on the mean HOLD cost at level $0.05/4$ stays below $U$.
When the rule of Table~\ref{tab:latency} fails, most of its answers miss exactly one of the ten members, at 0.7\% and 0.4\% more total distance on average.

\paragraph{Query shift and index updates}
\key{When a query shift or an index update breaks the risk target, recalibration on new CAL queries restores it at every level where the chain returns a rule.}
The guarantees hold for queries drawn like the samples, and text queries can follow another distribution~\cite{chen2024roargraph}.

\begin{figure}[t]
\centering
\includegraphics[width=\columnwidth]{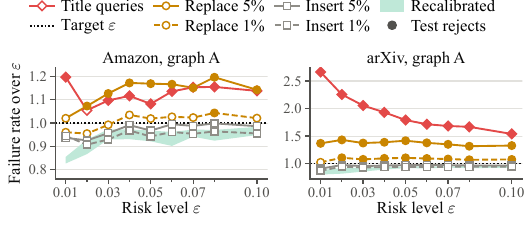}
\caption{Failure rate over the target on EVAL of graph A after a change of the queries or the index. Lines are the frontier rule calibrated before the change, markers filled where an exact binomial test at level $0.05/9$ rejects the target. The band is the range over the five changes after recalibration, which returns a rule except for title queries at 0.01 and keeps all within the target.}
\label{fig:shift}
\end{figure}

In Fig.~\ref{fig:shift}, we query graph A with only the titles of the CAL items and of 40{,}000 EVAL items, while the base items embed their full text. We also insert 1\% or 5\% new items in hnswlib, or replace as many base items in place.
Title queries complete later, and replacement prunes the bottom layer, whose mean degree falls by 10\% and 25\%.
The rules calibrated before these changes thus stop too early, and a one-sided exact binomial test at level $0.05/9$ rejects the target at 43 of the 54 levels of the title queries and the two replacements. Inserts leave the rules within the target.

Recalibration uses 40{,}000 CAL queries labeled and logged after the change, and the chain returns no rule only for title queries at 0.01.
The guarantee needs only this step, since the test holds for any candidate sequence fixed before CAL is drawn.
The recalibrated rule spends 0.92 to 1.03 times the visits of a full refit on FIT, SEL and CAL samples logged on the updated index, so the old chain stays efficient.
Without labels, a Kolmogorov-Smirnov distance on the stopping checkpoints of 1{,}000 queries, thresholded for 5\% false alarms, flags 54\% to 100\% of the title and replacement batches.
It does not test the risk, so an alarm still calls for new labeled CAL queries.

The official queries of YFCC-10M give a second shift, where the constraint changes the answer of 1.6\% of the queries.
Rules calibrated on held-out queries keep the target at all five levels on the official EVAL queries at 2.0\% to 2.8\% higher cost.
With all samples from the official queries, the ratio is 1.177 at 0.05 and 1.084 at 0.10.

\paragraph{Sample size and labeling cost}
\key{Labels are cheap, and a quarter of the CAL queries costs at most 5.6\% more visits.}
Per 10{,}000 sampled queries over one million items, the exact answers and logged trajectories take under 5 seconds on one GPU and 50 to 90 seconds on one 32-core server.
With 10{,}000 instead of 40{,}000 CAL queries, the rule at $\varepsilon=0.05$ spends 2.3\% to 5.6\% more visits, and with 30{,}000 instead of 120{,}000 EVAL queries its ratio grows by 0.036 to 0.044.
A deployment that needs no bound uses only FIT, SEL and CAL, 62{,}500 queries.

\subsubsection{Ablation study}

\paragraph{Richer learned policies}
\key{The strongest richer policy, the fully fitted one, is at most 0.5\% cheaper than our rule and at most 5.4\% costlier, with a median of 1.6\% more over four settings and seven risk levels.}
These policies and ours fit the same SEL queries, are calibrated on one half of CAL at level $\delta/M$ and are compared on the other half.
Per setting, the cross-fitted and declared-state policies cost a median of 1.6\% to 3.5\% and 1.2\% to 2.8\% more.
The advantage of the class is consistent with sample efficiency.
When SEL shrinks from 20{,}000 to 5{,}000 queries, the out-of-sample Lagrangian value of the class changes by at most 1\%. Over the same change, the gap of the cross-fitted policy to the class in this value, before any calibration, grows from between 5.6\% and 7.2\% to between 18.6\% and 26.7\%.

\paragraph{Interface and lower bound}
\key{The results do not hinge on a tuned interface, and bounding all risk levels at once loses at most 0.4\%.}
On graph A we vary the three parameters of the interface, with 40, 60 or 200 buckets, $J=50$ or 25, or a logistic-regression score.
The rule calibrated at $\varepsilon=0.05$ then costs between 0.8\% less and 2.5\% more than with the main interface, and the ratio stays at most 1.134 at every level from 0.03 to 0.10.
With the checkpoint and the flag alone, the empirical class optimum at 0.05 is 58\% and 51\% higher on Amazon and arXiv, yet its bound of 1.098 and 1.099 is about as small as that of the main interface.
The bound therefore measures what remains within a class, and only the cost ranks classes.
From 0.03 to 0.10, a lower bound for one risk level only lies within 0.4\% of the Lagrangian bound for all levels.

\section{Conclusion}
Declaring the interface of a stopping rule turns early termination into a problem with structure.
Over a monotone class, risk-constrained stopping is a linear program over an order polytope, and its Lagrangian minimizers form a nested chain of staircases that dynamic programming computes exactly.
This structure yields both a calibrated rule and a lower bound on the class optimum, and their ratio is a data-dependent optimality-gap bound within the declared class.
In our experiments, the rule replayed on logged searches stays within 1.132 of the class optimum at the risk levels 0.03 to 0.08 and 0.10, at confidence 95\%.
Larger savings thus need a rule outside the class.
At a common risk target, our rule also has a lower mean latency than every other calibrated rule we time.
We measure in-memory search on one machine with unfiltered traversal and fixed $k$ and capacities. The guarantees hold for queries drawn like the samples, so a query shift or an index update calls for recalibration.

\bibliographystyle{IEEEtran}
\bibliography{refs}

\begin{thebibliography}{10}
\providecommand{\url}[1]{#1}
\csname url@samestyle\endcsname
\providecommand{\newblock}{\relax}
\providecommand{\bibinfo}[2]{#2}
\providecommand{\BIBentrySTDinterwordspacing}{\spaceskip=0pt\relax}
\providecommand{\BIBentryALTinterwordstretchfactor}{4}
\providecommand{\BIBentryALTinterwordspacing}{\spaceskip=\fontdimen2\font plus
\BIBentryALTinterwordstretchfactor\fontdimen3\font minus
  \fontdimen4\font\relax}
\providecommand{\BIBforeignlanguage}[2]{{%
\expandafter\ifx\csname l@#1\endcsname\relax
\typeout{** WARNING: IEEEtran.bst: No hyphenation pattern has been}%
\typeout{** loaded for the language `#1'. Using the pattern for}%
\typeout{** the default language instead.}%
\else
\language=\csname l@#1\endcsname
\fi
#2}}
\providecommand{\BIBdecl}{\relax}
\BIBdecl

\bibitem{malkov2020}
Y.~A. Malkov and D.~A. Yashunin, ``Efficient and robust approximate nearest
  neighbor search using hierarchical navigable small world graphs,'' \emph{IEEE
  Transactions on Pattern Analysis and Machine Intelligence}, vol.~42, no.~4,
  pp. 824--836, 2020.

\bibitem{subramanya2019}
S.~{Jayaram Subramanya}, F.~Devvrit, H.~V. Simhadri, R.~Krishnaswamy, and
  R.~Kadekodi, ``{DiskANN}: Fast accurate billion-point nearest neighbor search
  on a single node,'' in \emph{Advances in Neural Information Processing
  Systems (NeurIPS)}, vol.~32, 2019.

\bibitem{sonmez2022}
T.~S{\"o}nmez and M.~B. Yenmez, ``Affirmative action in {India} via vertical,
  horizontal, and overlapping reservations,'' \emph{Econometrica}, vol.~90,
  no.~3, pp. 1143--1176, 2022.

\bibitem{ando2026}
S.~Ando and Y.~Yokoi, ``Ranking and rank aggregation with matroid prefix
  constraints,'' arXiv:2607.07153v2, 2026, to appear in ISAAC 2026.

\bibitem{li2020laet}
C.~Li, M.~Zhang, D.~G. Andersen, and Y.~He, ``Improving approximate nearest
  neighbor search through learned adaptive early termination,'' in
  \emph{Proceedings of the 2020 ACM SIGMOD International Conference on
  Management of Data}, 2020, pp. 2539--2554.

\bibitem{chatzakis2025darth}
M.~Chatzakis, Y.~Papakonstantinou, and T.~Palpanas, ``{DARTH}: Declarative
  recall through early termination for approximate nearest neighbor search,''
  \emph{Proceedings of the ACM on Management of Data}, vol.~3, no.~4, pp.
  1--26, 2025.

\bibitem{chatzakis2026darthplus}
------, ``{DARTH+}: Approximate nearest neighbor search with declarative recall
  and quality guarantees,'' HAL preprint hal-05566027, 2026.

\bibitem{horchidan2025conann}
S.~Horchidan, F.~Zeiher, H.~Bostr{\"o}m, and P.~Carbone, ``{ConANN}: Conformal
  approximate nearest neighbor search,'' \emph{Proceedings of the VLDB
  Endowment}, vol.~19, no.~1, pp. 29--42, 2025.

\bibitem{angelopoulos2021ltt}
A.~N. Angelopoulos, S.~Bates, E.~J. Cand{\`e}s, M.~I. Jordan, and L.~Lei,
  ``Learn then test: Calibrating predictive algorithms to achieve risk
  control,'' \emph{The Annals of Applied Statistics}, vol.~19, no.~2, pp.
  1641--1662, 2025.

\bibitem{schuster2022calm}
T.~Schuster, A.~Fisch, J.~Gupta, M.~Dehghani, D.~Bahri, V.~Tran, Y.~Tay, and
  D.~Metzler, ``Confident adaptive language modeling,'' in \emph{Advances in
  Neural Information Processing Systems (NeurIPS)}, vol.~35, 2022, pp.
  17\,456--17\,472.

\bibitem{jazbec2024fastyetsafe}
M.~Jazbec, A.~Timans, T.~H. Veljkovi{\'c}, K.~Sakmann, D.~Zhang, C.~A.
  Naesseth, and E.~Nalisnick, ``Fast yet safe: Early-exiting with risk
  control,'' in \emph{Advances in Neural Information Processing Systems
  (NeurIPS)}, vol.~37, 2024, pp. 129\,825--129\,854.

\bibitem{yang2021tao}
K.~Yang, H.~Wang, B.~Xu, W.~Wang, Y.~Xiao, M.~Du, and J.~Zhou, ``Tao: A
  learning framework for adaptive nearest neighbor search using static features
  only,'' arXiv:2110.00696, 2021.

\bibitem{zhang2026adaef}
C.~Zhang and R.~J. Miller, ``Distribution-aware exploration for adaptive {HNSW}
  search,'' \emph{Proceedings of the ACM on Management of Data}, vol.~4, no.~1,
  pp. 1--27, 2026.

\bibitem{teofili2025patience}
T.~Teofili and J.~Lin, ``Patience in proximity: A simple early termination
  strategy for {HNSW} graph traversal in approximate k-nearest neighbor
  search,'' in \emph{Advances in Information Retrieval, 47th European
  Conference on Information Retrieval (ECIR)}, ser. Lecture Notes in Computer
  Science, vol. 15574, 2025, pp. 401--407.

\bibitem{lu2026terminus}
J.~Lu, A.~Cidon, and M.~J. Freedman, ``When enough is enough: Rank-aware early
  termination for vector search,'' in \emph{Proceedings of Machine Learning and
  Systems (MLSys)}, 2026, oral presentation.

\bibitem{xia2026e2e}
W.~Xia, M.~Yang, W.~Li, and W.~Wang, ``{E2E}: Efficient filtered {AKNN} search
  via adaptive termination,'' in \emph{Proceedings of the 32nd ACM SIGKDD
  Conference on Knowledge Discovery and Data Mining V.2 (KDD)}, 2026, pp.
  5602--5613.

\bibitem{aljazzazi2025adaptive}
Y.~Al-Jazzazi, H.~Diwan, J.~Gou, C.~Musco, C.~Musco, and T.~Suel, ``Distance
  adaptive beam search for provably accurate graph-based nearest neighbor
  search,'' in \emph{Advances in Neural Information Processing Systems
  (NeurIPS)}, vol.~38, 2025, pp. 110\,958--110\,987.

\bibitem{angelopoulos2024crc}
A.~N. Angelopoulos, S.~Bates, A.~Fisch, L.~Lei, and T.~Schuster, ``Conformal
  risk control,'' in \emph{The Twelfth International Conference on Learning
  Representations (ICLR)}, 2024.

\bibitem{li2026spanner}
M.~Li, R.~Mittal, S.~Rana, S.~Shetiya, G.~Das, and N.~Koudas, ``{HNSW} with
  accuracy guarantees using graph spanners,'' arXiv:2607.02338, 2026.

\bibitem{zhang2023auncel}
Z.~Zhang, C.~Jin, L.~Tang, X.~Liu, and X.~Jin, ``Fast, approximate vector
  queries on very large unstructured datasets,'' in \emph{20th USENIX Symposium
  on Networked Systems Design and Implementation (NSDI)}, 2023, pp. 995--1011.

\bibitem{gao2023adsampling}
J.~Gao and C.~Long, ``High-dimensional approximate nearest neighbor search:
  with reliable and efficient distance comparison operations,''
  \emph{Proceedings of the ACM on Management of Data}, vol.~1, no.~2, pp.
  137:1--137:27, 2023.

\bibitem{bates2021rcps}
S.~Bates, A.~Angelopoulos, L.~Lei, J.~Malik, and M.~I. Jordan,
  ``Distribution-free, risk-controlling prediction sets,'' \emph{Journal of the
  ACM}, vol.~68, no.~6, pp. 43:1--43:34, 2021.

\bibitem{luedtke2008}
J.~Luedtke and S.~Ahmed, ``A sample approximation approach for optimization
  with probabilistic constraints,'' \emph{SIAM Journal on Optimization},
  vol.~19, no.~2, pp. 674--699, 2008.

\bibitem{wang2008saa}
W.~Wang and S.~Ahmed, ``Sample average approximation of expected value
  constrained stochastic programs,'' \emph{Operations Research Letters},
  vol.~36, no.~5, pp. 515--519, 2008.

\bibitem{mak1999}
W.-K. Mak, D.~P. Morton, and R.~K. Wood, ``Monte {Carlo} bounding techniques
  for determining solution quality in stochastic programs,'' \emph{Operations
  Research Letters}, vol.~24, no. 1--2, pp. 47--56, 1999.

\bibitem{vieira2011diversification}
M.~R. Vieira, H.~L. Razente, M.~C.~N. Barioni, M.~Hadjieleftheriou,
  D.~Srivastava, C.~Traina, and V.~J. Tsotras, ``On query result
  diversification,'' in \emph{Proceedings of the 27th IEEE International
  Conference on Data Engineering}, 2011, pp. 1163--1174.

\bibitem{anand2025}
P.~Anand, P.~Indyk, R.~Krishnaswamy, S.~Mahabadi, V.~C. Raykar, K.~Shiragur,
  and H.~Xu, ``Graph-based algorithms for diverse similarity search,'' in
  \emph{Proceedings of the 42nd International Conference on Machine Learning
  (ICML)}, ser. Proceedings of Machine Learning Research, vol. 267, 2025, pp.
  1483--1504.

\bibitem{chen2026}
J.~Chen, S.~D. Nagawanshi, S.~Xie, H.~Xu, A.~Zhou, and S.~Zhou, ``Fast diverse
  nearest neighbor search,'' in \emph{Advances in Neural Information Processing
  Systems (NeurIPS), poster}, 2026.

\bibitem{celis2018}
L.~E. Celis, D.~Straszak, and N.~K. Vishnoi, ``Ranking with fairness
  constraints,'' in \emph{45th International Colloquium on Automata, Languages,
  and Programming (ICALP)}, ser. LIPIcs, vol. 107, 2018, pp. 28:1--28:15.

\bibitem{gollapudi2023filtered}
S.~Gollapudi, N.~Karia, V.~Sivashankar, R.~Krishnaswamy, N.~Begwani, S.~Raz,
  Y.~Lin, Y.~Zhang, N.~Mahapatro, P.~Srinivasan, A.~Singh, and H.~V. Simhadri,
  ``Filtered-{DiskANN}: Graph algorithms for approximate nearest neighbor
  search with filters,'' in \emph{Proceedings of the ACM Web Conference 2023
  (WWW)}, 2023, pp. 3406--3416.

\bibitem{wang2023nhq}
M.~Wang, L.~Lv, X.~Xu, Y.~Wang, Q.~Yue, and J.~Ni, ``An efficient and robust
  framework for approximate nearest neighbor search with attribute
  constraint,'' in \emph{Advances in Neural Information Processing Systems
  (NeurIPS)}, vol.~36, 2023, pp. 15\,738--15\,751.

\bibitem{patel2024acorn}
L.~Patel, P.~Kraft, C.~Guestrin, and M.~Zaharia, ``{ACORN}: Performant and
  predicate-agnostic search over vector embeddings and structured data,''
  \emph{Proceedings of the ACM on Management of Data}, vol.~2, no.~3, pp.
  1--27, 2024.

\bibitem{fagin2003}
R.~Fagin, A.~Lotem, and M.~Naor, ``Optimal aggregation algorithms for
  middleware,'' \emph{Journal of Computer and System Sciences}, vol.~66, no.~4,
  pp. 614--656, 2003.

\bibitem{theobald2004topk}
M.~Theobald, G.~Weikum, and R.~Schenkel, ``Top-k query evaluation with
  probabilistic guarantees,'' in \emph{Proceedings of the 30th International
  Conference on Very Large Data Bases (VLDB)}, 2004, pp. 648--659.

\bibitem{ciaccia2000pac}
P.~Ciaccia and M.~Patella, ``{PAC} nearest neighbor queries: Approximate and
  controlled search in high-dimensional and metric spaces,'' in
  \emph{Proceedings of the 16th International Conference on Data Engineering},
  2000, pp. 244--255.

\bibitem{rogers2002}
L.~C.~G. Rogers, ``Monte {Carlo} valuation of {American} options,''
  \emph{Mathematical Finance}, vol.~12, no.~3, pp. 271--286, 2002.

\bibitem{haugh2004}
M.~B. Haugh and L.~Kogan, ``Pricing {American} options: A duality approach,''
  \emph{Operations Research}, vol.~52, no.~2, pp. 258--270, 2004.

\bibitem{oxley2011}
J.~Oxley, \emph{Matroid Theory}, 2nd~ed.\hskip 1em plus 0.5em minus 0.4em\relax
  Oxford University Press, 2011.

\bibitem{stanley1986}
R.~P. Stanley, ``Two poset polytopes,'' \emph{Discrete \& Computational
  Geometry}, vol.~1, no.~1, pp. 9--23, 1986.

\bibitem{macmahon1916}
P.~A. MacMahon, \emph{Combinatory Analysis}.\hskip 1em plus 0.5em minus
  0.4em\relax Cambridge University Press, 1916, vol.~II.

\bibitem{topkis1978}
D.~M. Topkis, ``Minimizing a submodular function on a lattice,''
  \emph{Operations Research}, vol.~26, no.~2, pp. 305--321, 1978.

\bibitem{picard1976}
J.-C. Picard, ``Maximal closure of a graph and applications to combinatorial
  problems,'' \emph{Management Science}, vol.~22, no.~11, pp. 1268--1272, 1976.

\bibitem{gallo1989}
G.~Gallo, M.~D. Grigoriadis, and R.~E. Tarjan, ``A fast parametric maximum flow
  algorithm and applications,'' \emph{SIAM Journal on Computing}, vol.~18,
  no.~1, pp. 30--55, 1989.

\bibitem{barlow1972}
R.~E. Barlow, D.~J. Bartholomew, J.~M. Bremner, and H.~D. Brunk,
  \emph{Statistical Inference under Order Restrictions}.\hskip 1em plus 0.5em
  minus 0.4em\relax John Wiley \& Sons, 1972.

\bibitem{maurer2009}
A.~Maurer and M.~Pontil, ``Empirical {Bernstein} bounds and sample variance
  penalization,'' in \emph{Proceedings of the 22nd Annual Conference on
  Learning Theory (COLT)}, 2009.

\bibitem{hou2026}
Y.~Hou, J.~Li, X.~Fu, Z.~He, A.~Yan, X.~Chen, and J.~McAuley, ``Bridging
  language and items for retrieval and recommendation: Benchmarking {LLMs} as
  semantic encoders,'' in \emph{Proceedings of the 64th Annual Meeting of the
  Association for Computational Linguistics (Volume 1: Long Papers)}, 2026, pp.
  3251--3265.

\bibitem{reimers2019}
N.~Reimers and I.~Gurevych, ``{Sentence-BERT}: Sentence embeddings using
  {Siamese} {BERT}-networks,'' in \emph{Proceedings of the 2019 Conference on
  Empirical Methods in Natural Language Processing and the 9th International
  Joint Conference on Natural Language Processing (EMNLP-IJCNLP)}, 2019, pp.
  3980--3990.

\bibitem{simhadri2025bigann}
H.~V. Simhadri, M.~Aum{\"u}ller, M.~Douze, D.~Baranchuk, A.~Ingber, E.~Liberty,
  G.~Williams, B.~Landrum, M.~Manohar, M.~Karjikar, L.~Dhulipala, M.~Chen,
  Y.~Chen, R.~Ma, K.~Zhang, Y.~Cai, J.~Shi, W.~Zheng, Y.~Chen, J.~Yin, and
  B.~Huang, ``Results of the {Big ANN}: {NeurIPS}'23 competition,'' in
  \emph{Advances in Neural Information Processing Systems 38 (NeurIPS 2025),
  Datasets and Benchmarks Track}, 2025.

\bibitem{radford2021clip}
A.~Radford, J.~W. Kim, C.~Hallacy, A.~Ramesh, G.~Goh, S.~Agarwal, G.~Sastry,
  A.~Askell, P.~Mishkin, J.~Clark, G.~Krueger, and I.~Sutskever, ``Learning
  transferable visual models from natural language supervision,'' in
  \emph{Proceedings of the 38th International Conference on Machine Learning
  (ICML)}, ser. Proceedings of Machine Learning Research, vol. 139, 2021, pp.
  8748--8763.

\bibitem{thomee2016yfcc}
B.~Thomee, D.~A. Shamma, G.~Friedland, B.~Elizalde, K.~Ni, D.~Poland, D.~Borth,
  and L.-J. Li, ``{YFCC100M}: The new data in multimedia research,''
  \emph{Communications of the ACM}, vol.~59, no.~2, pp. 64--73, 2016.

\bibitem{abuelhaija2016yt8m}
S.~Abu-El-Haija, N.~Kothari, J.~Lee, P.~Natsev, G.~Toderici, B.~Varadarajan,
  and S.~Vijayanarasimhan, ``{YouTube-8M}: A large-scale video classification
  benchmark,'' arXiv:1609.08675, 2016.

\bibitem{longstaff2001}
F.~A. Longstaff and E.~S. Schwartz, ``Valuing {American} options by simulation:
  A simple least-squares approach,'' \emph{Review of Financial Studies},
  vol.~14, no.~1, pp. 113--147, 2001.

\bibitem{chen2024roargraph}
M.~Chen, K.~Zhang, Z.~He, Y.~Jing, and X.~S. Wang, ``{RoarGraph}: A projected
  bipartite graph for efficient cross-modal approximate nearest neighbor
  search,'' \emph{Proceedings of the VLDB Endowment}, vol.~17, no.~11, pp.
  2735--2749, 2024.

\end{thebibliography}

\end{document}